\documentclass[12pt,preprint]{aastex}

\newcommand\wcm{W~cm$^{-2}$~\micron$^{-1}$}
\newcommand\mks{J~K$^{-1}$~m$^{-2}$~s$^{-1/2}$}
\begin{document}

\title{A Spitzer Study of Comets 2P/Encke, 67P/Churyumov-Gerasimenko,
  and C/2001 HT50 (LINEAR-NEAT)}

\shorttitle{A Spitzer Study of Three Comets}

\author{Michael S. Kelley,\altaffilmark{1}
Charles E. Woodward,\altaffilmark{1}
David E. Harker,\altaffilmark{2}
Diane H. Wooden,\altaffilmark{3}
Robert D. Gehrz,\altaffilmark{1}
Humberto Campins,\altaffilmark{4}
Martha S. Hanner,\altaffilmark{5}
Susan M. Lederer,\altaffilmark{6}
David J. Osip,\altaffilmark{7}
Jana Pittichov{\'a},\altaffilmark{8,9}
and Elisha Polomski\altaffilmark{1}}

\email{msk@astro.umn.edu}

\altaffiltext{1}{Department of Astronomy, University of Minnesota, 116
Church St SE, Minneapolis, MN 55455}

\altaffiltext{2}{Center for Astronomy and Space Science, University of
California, San Diego, 9500 Gilman Drive, Department 0424, San Diego,
CA 92093}

\altaffiltext{3}{NASA Ames Research Center, Space Science Division,
MS 245-3, Moffett Field, CA 94035}

\altaffiltext{4}{Department of Physics, University of Central Florida,
Orlando, FL 32816}

\altaffiltext{5}{Department of Astronomy, University of Massachusetts,
710 North Pleasant Street, Amherst, MA 01003}

\altaffiltext{6}{Department of Physics, California State University,
5500 University Parkway, San Bernardino, CA 92407}

\altaffiltext{7}{Las Campanas Observatory, Carnegie Observatories,
Casilla 601, La Serena, Chile}

\altaffiltext{8}{Institute for Astronomy, University of Hawaii at
Manoa, 2680 Woodlawn Drive, Honolulu, HI 96822}

\altaffiltext{9}{Astronomical Institute, Slovak Academy of Sciences,
Dubravska cesta 9, 845 04 Bratislava, Slovak Republic}

\begin{abstract}
We present infrared images and spectra of comets 2P/Encke,
67P/Churyumov-Gerasimenko, and C/2001~HT50 (LINEAR-NEAT) as part of a
larger program to observe comets inside of 5~AU from the sun with the
\textit{Spitzer Space Telescope}.  The nucleus of comet 2P/Encke was
observed at two vastly different phase angles (20\degr{} and
63\degr).  Model fits to the spectral energy distributions of the
nucleus suggest comet Encke's infrared beaming parameter derived from
the near-Earth asteroid thermal model may have a phase angle
dependence.  The observed emission from comet Encke's dust coma is
best-modeled using predominately amorphous carbon grains with a grain
size distribution that peaks near 0.4~\micron{}, and the silicate
contribution by mass to the sub-micron dust coma is constrained to
$<31$\%.  Comet 67P/Churyumov-Gerasimenko was observed with distinct
coma emission in excess of a model nucleus at a heliocentric distance
of 5.0~AU.  The coma detection suggests that sublimation processes are
still active or grains from recent activity remain near the nucleus.
Comet C/2001~HT50 (LINEAR-NEAT) showed evidence for crystalline
silicates in the spectrum obtained at 3.2~AU and we derive a
silicate-to-carbon dust ratio of 0.6.  The ratio is an order of
magnitude lower than that derived for comets 9P/Tempel~1 during the
\textit{Deep Impact} encounter and C/1995~O1 (Hale-Bopp).
\end{abstract}


\section{INTRODUCTION}
Comets are frozen reservoirs of primitive solar dust grains and ices.
Analysis of the composition and size distribution of cometary dust
grains from infrared imaging and spectroscopic observations expedites
an appraisal of the physical characteristics of the solid materials
that constituted the primitive solar nebula \citep{ahearn04, ehrenf04,
wooden05}. The study of comets is an indirect probe of the origin of
the constituents of the primitive solar system, their subsequent
evolution into planetesimals, and their relationship to materials in
other astrophysical environments \citep{wooden05}.

Although comets of all types have undergone some amount of
post-formation processing, they remain the best preserved sources of
material extant during our solar system's epoch of planet formation.
In the current paradigm, the nearly isotropic comets (i.e., Oort cloud
and Halley-type comets) formed amongst the giant planets and were
scattered into large orbits, in a spherically symmetric manner
\citep{dones04}.  The ecliptic comets (including Jupiter-family
comets) originate from the Kuiper-belt and scattered disk populations
and likely formed \textit{in situ} or in the transneptunian region
\citep{duncan04, morbidelli04}.  Comparisons between the nearly
isotropic and ecliptic comets may reveal the differences in their
post-formation processing or the structure and mineralogy of the
proto-planetary disk.  Both nearly isotropic and ecliptic comets have
been exposed to bombardment by UV photons and cosmic rays, although to
varying extents (the Jupiter-family comets have been exposed to
$\sim10^5$ times more UV and 100~KeV solar cosmic rays than the Oort
cloud comets).  The ecliptic comets have suffered frequent collisions
during their residence in the Kuiper-belt and are likely to be
fragments of larger Kuiper-belt bodies \citep{stern03}.  The number of
comets studied by mid-infrared spectroscopic methods necessary to
determine their detailed mineralogies is increasing \citep[e.g.,
see][]{hanner96, harker06, harker05, harker02, harker99, honda04,
kelley05b, lynch02, lynch00, sitko04, wooden04} and we may soon be
able to compare comets to each other as groups, rather than
individually.

We present \textit{Spitzer Space Telescope} \citep{werner04} images
and spectra of comets 2P/Encke, a 3.3~yr period ($P$), Jupiter-family
comet with a perihelion distance, $q=0.3$~AU, known for an abundance
of large dust particles \citep{reach00} and an extensive debris trail
\citep{sykes92}; 67P/Churyumov-Gerasimenko (67P), $P=6.6$~yr,
$q=1.3$~AU, a Jupiter-family comet and the primary mission target of
the European Space Agency's \textit{Rosetta} spacecraft; and
C/2001~HT50 (LINEAR-NEAT) (HT50), a long period, Oort cloud comet,
$P=40,250$~yr, $q=2.8$~AU.

Comet Encke frequently approaches the Earth on it's 3.3~yr orbit and
is one of the most studied of all comets \citep{sekanina91}.  Comet
Encke was also one of the first comets discovered to have a dust trail
\citep{sykes92}.  Dust trails are composed of large ($\gtrsim0.1$~mm),
slow moving particles and are precursors to meteor streams [Encke is
associated with the Taurid meteor stream \citep{whipple50}].  The
comet also exhibits weak or non-existent 10~\micron{} silicate
emission \citep{campins82, gehrz89, lisse04}.  The existence of a dust
trail, association with a meteor stream, and the lack of a strong
silicate feature has led investigators to conclude that Encke's dust
production is dominated by large particles.  We present
\textit{Spitzer Space Telescope} observations of comet Encke in
\S\ref{spectra-obs-text}, and derive the comet's dust coma mineralogy
at 2.4~AU in \S\ref{encke-coma}.  The mineralogy of comet Encke is
discussed in \S\ref{mineral-discuss}.  We derive the temperature and
effective size of the nucleus of comet Encke in \S\ref{encke-results}
and discuss the results in \S\ref{eta-discuss}.

Comet 67P is the primary target of the \textit{Rosetta} mission.  The
spacecraft is designed to intercept and orbit the comet at $r_h =
4.5$~AU (pre-perihelion) to study the development of coma activity as
the comet approaches the
sun\footnote{\url{http://www.esa.int/esaMI/Rosetta/}}.  Information on
the comet's dust environment is crucial to mission planning, which
motivated our \textit{Spitzer} observations of the comet at 5.0~AU
(post-perihelion).  The observation is presented in
\S\ref{spectra-obs-text} and the results presented in
\S\ref{cg-results}.

Comet HT50 is an Oort cloud comet with an orbital period that suggests
it has orbited the sun many times ($P=40,250$~yr).  It was discovered
to be cometary at the large heliocentric distance of 7.5~AU
\citep{pravdo01}.  We observed comet HT50 twice with \textit{Spitzer}.
Both observations are presented in \S\ref{spectra-obs-text} and we
derive dust mineralogies at both epochs in \S\ref{ht50-results}.  We
discuss HT50's mineral content and compare it to other Oort cloud
comets in \S\ref{mineral-discuss}.

\section{OBSERVATIONS AND REDUCTION}
\subsection{Spectra}\label{spectra-obs-text}
Spectra of comets Encke, 67P, and HT50 were obtained with the Infrared
Spectrograph (IRS) instrument \citep{houck04} on \textit{Spitzer}.
All comets were observed in both low- ($R \approx 64$--128) and high-
($R\approx600$) resolution modes, although not at all wavelengths.
The slit widths, aperture extraction sizes (described below), and slit
orientations are listed in Table~\ref{apertures}.  Note that the slit
parameters force us to sample different areas of each comet's coma
(even though each slit extraction is centered on the nucleus) and
remain as a source of unknown error in our resultant spectra.  A
summary of our observations is presented in Table~\ref{spec-log}.
There were two observations of comet Encke, one from our
\textit{Spitzer} guaranteed time observation (GTO) program, program
identification (PID)~119, and one from PID~210, one of comet 67P from
our general observer proposal (PID~2316), and two temporally distinct
observations of HT50 from our GTO program (PID~131).

Post-pipeline spectral reductions were performed on IRS pipeline
S12.0.2 basic calibrated data (BCD) frames.  Our reduction method
proceeded as follows: 1) The BCD frames were two-dimensionally
background subtracted (when possible, as described below) and fatally
bad pixels were removed by nearest-neighbor interpolation.  2)
Spectral extraction and initial calibration were performed by the
\textit{Spitzer} IRS Custom Extraction (SPICE) tool, available from
the SSC\footnote{Available at
\url{http://ssc.spitzer.caltech.edu/postbcd/}}.  If a two-dimensional
background subtraction was not possible, we subtracted a background
model from the extracted spectra at this step.  3) We removed the
nucleus contribution, if known, from the extracted spectra.  Nuclei
were modeled with the near-Earth asteroid thermal model
\citep{harris98, delbo02} as described in \S\ref{encke-results}. 4) We
corrected the combined, nucleus subtracted spectra with our derived
extended source calibration (described below). 5) We scaled each
module to produce one continuous spectrum.  The scaling removed
module-to-module mis-calibrations and corrected for differences in
module slit orientations with respect to the comet comae.  6) Finally,
we scaled the entire spectrum to agree with aperture photometry
derived from \textit{Spitzer} imaging (\S\ref{imaging}), if possible.

All spectra were extracted with a constant width aperture to produce a
final spectrum representative of a given aperture size around the
comet (Table~\ref{apertures}).  Our SL and LL apertures were
17.5\arcsec{} and 51.5\arcsec{} on the sky ($\approx 1/3$ of the
slit), i.e., they encompass the first diffraction ring of a point
source at the longest wavelength of each module.  The default point
source optimized aperture used by the IRS calibration pipeline has an
aperture size that varies with wavelength, i.e., the default aperture
width in arc seconds is proportional to wavelength.  Such an aperture
encompasses a diffraction limited point source (but a varying amount
of sky) at every extracted wavelength.  Therefore, the point source
optimized aperture is not optimized for extended source extractions.
We compared spectra of the IRS calibration star HD~173511 extracted
with our constant width apertures to spectra extracted with the
default point source tuned apertures.  The differences in the
point-to-point photometry (aperture averaged at a given wavelength)
between the two spectra were $\lesssim2$\%.  We note the IRS pipeline
uses the entire available slit by default in the high-resolution
modules.

\begin{figure}
\plotone{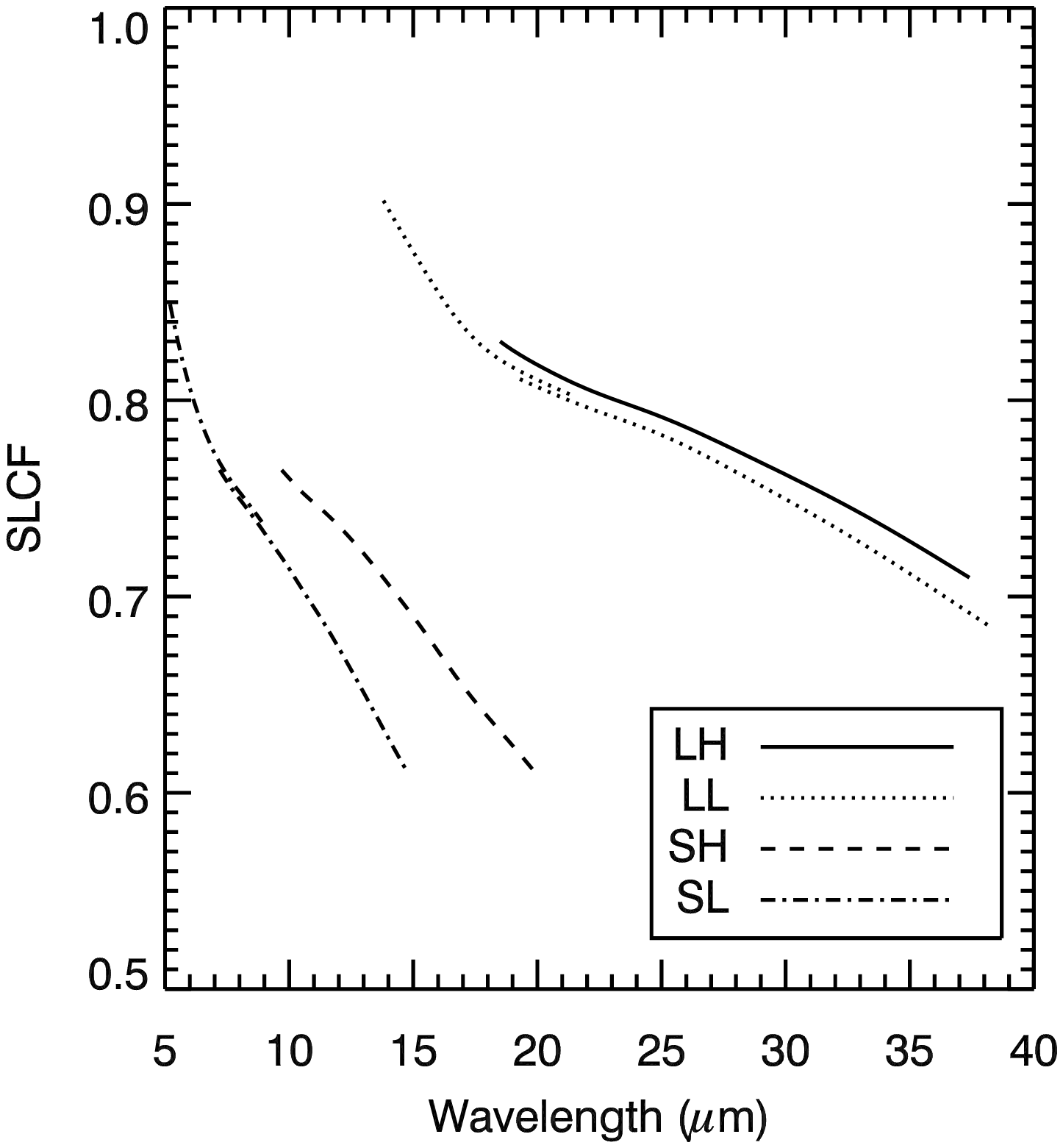}
\caption{Slit loss correction factor for all IRS modules (see
\S\ref{spectra-obs-text} for a discussion).}
\label{slcf}
\end{figure}

An automated spectral calibration of extended sources was not
available from the SSC, yet the extent of the comae of Encke and HT50
(described below) requires us to calibrate these spectra for extended
sources.  Currently, point sources are used for IRS calibration
targets.  We estimate the narrow entrance slits of the IRS modules
block up to 40\% of the flux of a point source at the longest
wavelengths.  Although this does not introduce additional absolute
photometric uncertainty into the final data products when calibrating
point sources (as this same fraction is always removed from all point
sources), the slit-loss does affect the photometric precision for
sources larger than the point spread function of the instrument. The
emission from comets is comprised of emission from the nucleus
(generally spatially unresolved) plus a contribution from the coma,
which is a region of radially and azimuthally varying surface
brightness.  To correctly flux calibrate the comet spectra, the
nucleus spectrum must be subtracted if the nucleus flux is a
significant fraction of the total emission in the slit (step~3,
above).  Inspection of the IRS peak-up images is used to confirm the
validity of this assumption for any individual comet (cf,
\S\ref{imaging}).  Comets Encke and 67P required nucleus subtraction,
HT50 did not.  The Encke nucleus is discussed in \S\ref{encke-results}
and the 67P nucleus is discussed in \S\ref{cg-results}.  In all cases,
we assume the coma surface brightness is a slowly varying function
over the slit width (again confirmed by the peak-up images).  Next, we
generate an image of a point source at each extracted wavelength for
an obscured primary with \textit{Spitzer}'s optical parameters
\citep{werner04}.  A slit mask is applied to the image and the flux in
the slit mask ($I_\lambda$) is compared to the total flux of the
simulated point source ($I_{\lambda,0}$).  The slit-loss correction
factor is $1 - I_\lambda/I_{\lambda,0}$ (Fig.~\ref{slcf}).  The
coefficients of third-order polynomial fits to the slit-loss
correction factors ($\lambda$ in units of \micron) for the four IRS
modules are given in Table~\ref{slcf-table}.  This slit-loss
correction factor is multiplicatively applied at each wavelength to
the nuclear flux subtracted spectra.

The five observations (Table~\ref{spec-log}) were designed with
various observing strategies.  The first observation of Encke
\dataset[ADS/Sa.Spitzer\#6582016]{(PID~119, AORKEY~6582016)} was
designed to obtain high-resolution spectra of the comet from
10--38~\micron{} and low-resolution spectra from 5--14~\micron{}.  It
was executed on 2004 June 25 at 05:28~UT when Encke was at a
heliocentric distance, $r_h$, of 2.573~AU, a \textit{Spitzer}-comet
distance, $\Delta_s$, of 1.985~AU, and a phase angle, $\phi$, of
21\degr.  The observation used an IRS spectral map astronomical
observing request (AOR).  We describe spectral map AORs with
$n_\parallel \times n_\perp$ notation, where $n_\parallel$ is the
number of steps parallel to the slit's long (spatial) direction, and
$n_\perp$ is the number of steps perpendicular to the long direction.
The first comet Encke observation used a $1 \times 3$ map of
60\arcsec{} steps perpendicular to the slit direction for all three
modules.  The two 60\arcsec{} steps off-source provide background
measurements for the on-source position.  A red peak-up image was used
to acquire comet Encke and center the IRS slits.

The second observation of Encke
\dataset[ADS/Sa.Spitzer\#6613248]{(PID~210, AORKEY~6613248)} only used
low-resolution modules to cover the entire 5--38~\micron{} spectral
region and was executed on 2004 June 25 at 08:27~UT ($r_h = 2.577$~AU,
$\Delta_s = 1.982$~AU, $\phi = 21$\degr). A blue peak-up image was
used to acquire comet Encke. The AOR for the 5--14~\micron{}
low-resolution module (SL) executed a $2 \times 1$ spectral map with
20\arcsec{} steps parallel to the slit and the 14--38~\micron{}
low-resolution module (LL) executed a $3 \times 1$ map with
40\arcsec{} steps parallel to the slit.  The parallel motion was
small enough such that the comet was located in the slit for each map
position.

\begin{figure}
\plotone{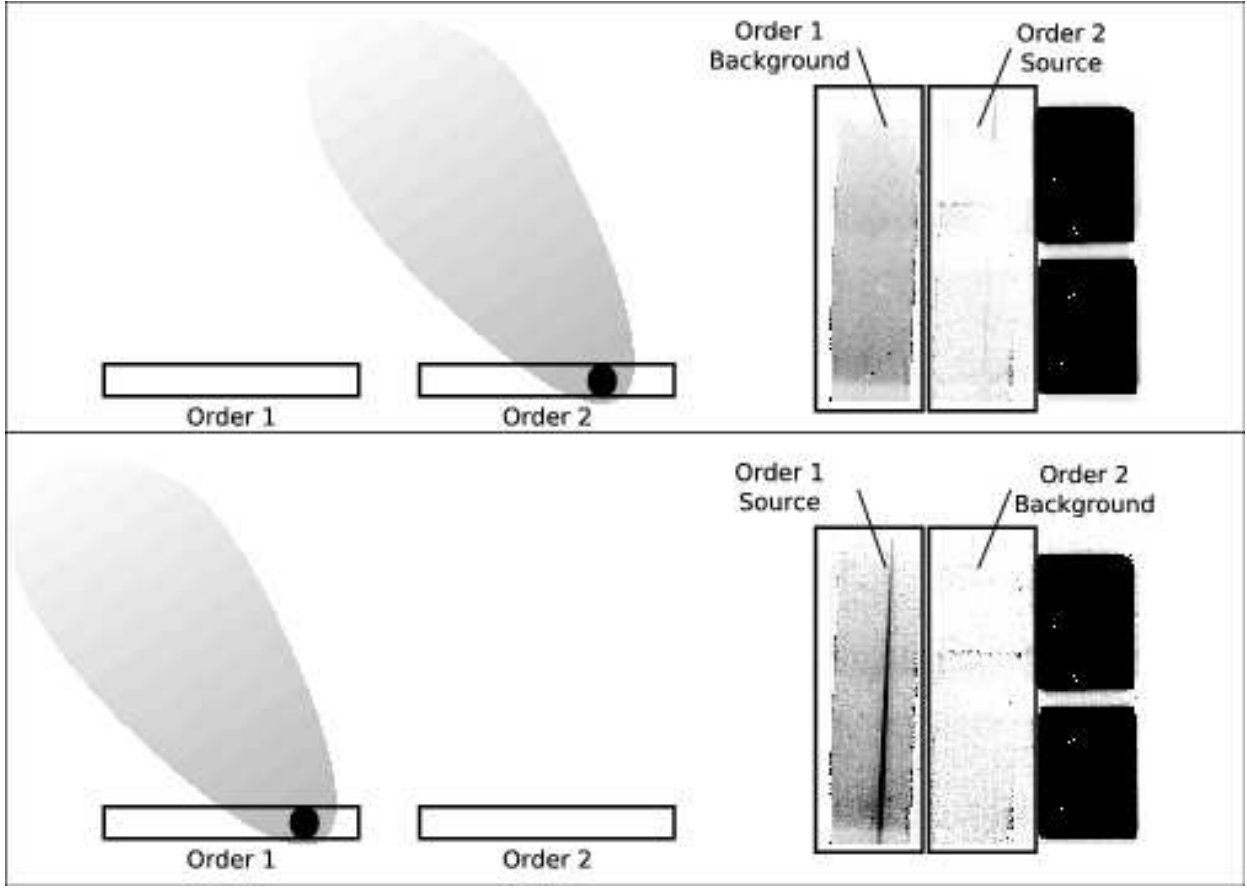}
\caption{Schematic illustrating how background is removed in
\textit{Spitzer}/IRS low-resolution observations of extended sources.
When the IRS observes the science target in one order (upper panel),
the accompanying order provides a measurement of the background (lower
panel).  This strategy allows the background for a particular order to
be two-dimensionally subtracted from the science observation.}
\label{bonus-background}
\end{figure}

Each low-resolution module consist of two slits, one of which is
centered on the science target and the other offset $\approx
60$\arcsec{} for SL and $\approx90$\arcsec{} for LL from the
comet (Fig.~\ref{bonus-background}).  We use the extra slit position
to background subtract our spectra.  For example, when the comet is in
the 1st order SL slit (SL1, 7--14~\micron), sky background is measured
in the 2nd order SL slit (SL2, 5--7~\micron).  The spacecraft is
nodded to place the comet in the SL2 slit, providing a measurement of
the sky background in SL1 (see Fig.~\ref{bonus-background}).  For the
second observation of comet Encke, the off-source order is used as a
background measurement and two-dimensionally subtracted from the
science order.

The observation of comet 67P
\dataset[ADS/Sa.Spitzer\#10204928]{(PID~2316, AORKEY~10204928)} used
an IRS spectral stare AOR in both long wavelength modules and resulted
in a spectrum spanning 14--38~\micron.  This AOR was executed on 2004
July 15 at 08:13~UT ($r_h = 4.987$~AU, $\Delta_s = 4.744$~AU,
$\phi=12$\degr).  The red peak-up array shows a point source at the
location of the comet.  To remove the background in LL, we subtracted
nod pairs and extracted the resulting spectra.  The high-resolution
module did not detect the comet and will no longer be discussed.

Two observations of comet HT50 were executed.  The first observation,
obtained on 2003 December 17 at 15:39~UT ($r_h = 3.238$~AU, $\Delta_s
= 2.652$~AU, $\phi=16$\degr), was an IRS spectral map AOR
\dataset[ADS/Sa.Spitzer\#6589440]{(PID~131, AORKEY~6589440)} of size
$1 \times 3$ with 7.2\arcsec{} steps ($\perp$) SL, a $2 \times 3$
map of $7.0\arcsec{} \times 4.8\arcsec{}$ ($\parallel \times \perp$)
for the short wavelength, high-resolution module (SH), and a $2 \times
3$ map of $12.8\arcsec{} \times 9.6\arcsec{}$ ($\parallel \times
\perp$) for the long wavelength, high-resolution module (LH).  We
extracted spectra from the map positions with the greatest amount of
signal.  The second observation
\dataset[ADS/Sa.Spitzer\#11625472]{(PID~131, AORKEY~11625472)} was
obtained on 2004 July 18 at 10:36~UT ($r_h = 4.598$~AU, $\Delta_s =
4.368$~AU, $\phi=13$\degr) with an IRS spectral stare AOR for both
high-resolution modules and the first order (7--14~\micron) of the SL
module.  There were no measurements of the background for the SH and
LH modules.  Instead, the background flux density was estimated using
the background model in the \textit{Spitzer Planning Observations
Tool} (SPOT) provided by the \textit{Spitzer} Science Center
(SSC)\footnote{Available at
\url{http://ssc.spitzer.caltech.edu/propkit/spot/}}.  The model
background was subtracted after spectral extraction.  Restricting the
second epoch SL observation of HT50 to a one order spectral stare does
not permit us to use the same two-dimensional background subtraction
techniques employed for comet Encke. Furthermore, the extent of the
coma in HT50 does not permit us to directly nod subtract the images to
remove the background as described above for comet 67P (67P was a
point source within the slit).  However, the SL module is long enough
in the spatial dimension ($\approx60$\arcsec{} long) such that the
coma of HT50 did not entirely fill the slit.  To determine the
background contribution in SL, a Gaussian plus linear background term
was fitted to the spatial dimension for every $\lambda_i$.  The linear
background term was subtracted from the extracted spectrum.

\subsection{Imaging}\label{imaging}

\begin{figure}
\plotone{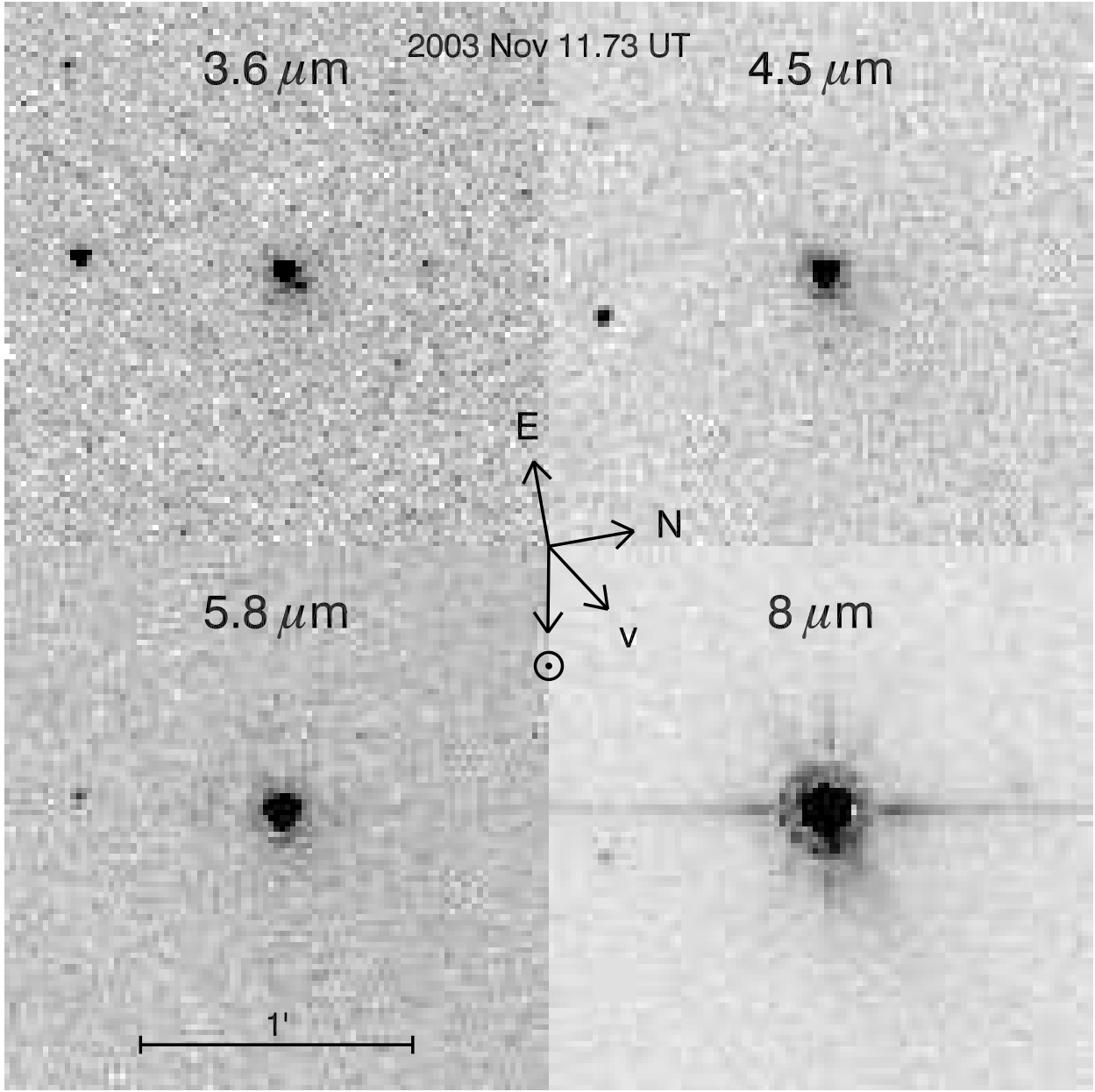}
\caption{IRAC images of comet 2P/Encke obtained on 2003 Nov 11.73 UT.
Each image displays a $2\arcmin{} \times 2\arcmin{}$ area centered on
the comet with linear intensity scale.  Arrows mark the image
orientation (N, E), the projected velocity vector (v), and sun
direction ($\sun$) as seen by \textit{Spitzer}.  The 8~\micron{} image
saturated on the nucleus.}
\label{nov-encke-images}
\end{figure}

\begin{figure}
\epsscale{0.6}
\plotone{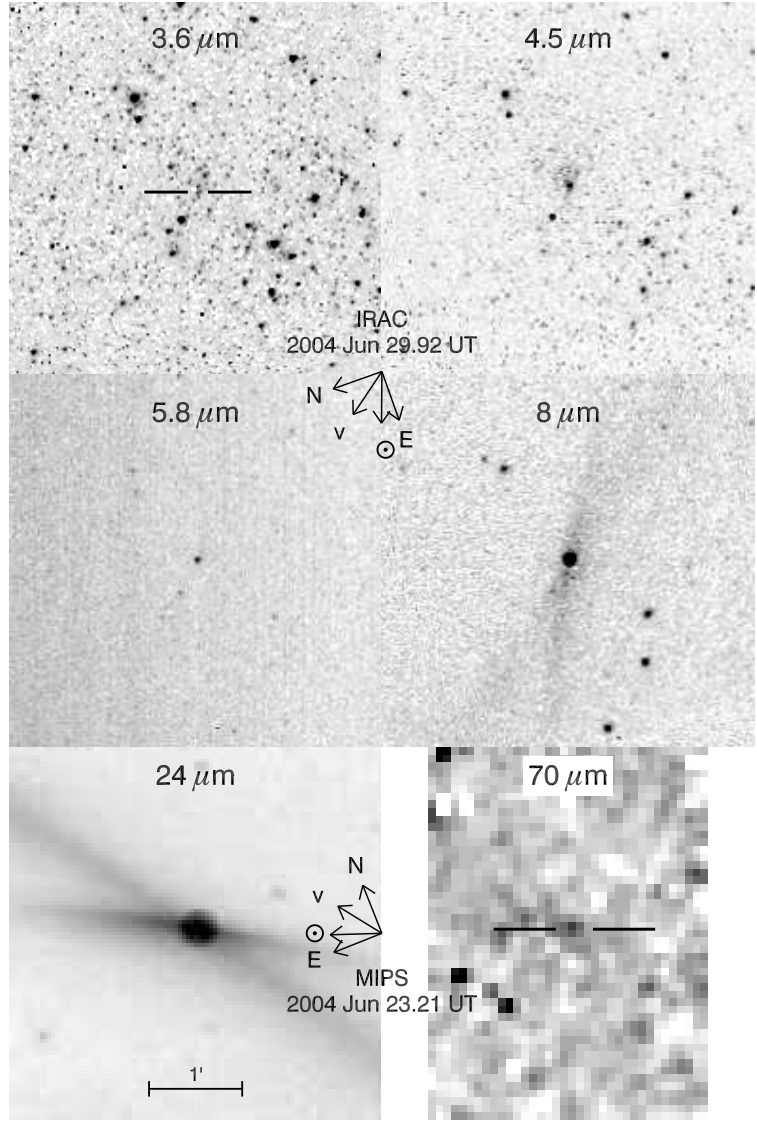}
\epsscale{1.0}
\caption{IRAC and MIPS mosaics of comet 2P/Encke obtained in 2004
June.  Each image displays a $4\arcmin{} \times 4\arcmin{}$ area
centered on the comet with a linear intensity scale.  When the comet
position is unclear, horizontal lines mark the location of the
nucleus.  The arrows are the same as in Fig.~\ref{nov-encke-images}.
The 3.6 to 8~\micron{} orientations are labeled by the top set of
arrows, and the 24 and 70~\micron{} orientations are labeled by the
bottom set of arrows.  The near horizontal ``spikes'' in the
24~\micron{} image are from recent comet activity and the emission
stretching diagonally across the image is the dust trail
\citep{gehrz06}.}
\label{jun-encke-images}
\end{figure}

All comets were imaged at the time of spectra acquisition by the
peak-up mode of the IRS.  Comet Encke was also observed with the
Infrared Array Camera (IRAC) in all band-passes during
\textit{Spitzer's} in-orbit checkout and science verification phase
(W.~T.\mbox{} Reach et al.~2006, in preparation) on 2003 November 11
at 17:33~UT ($r_h = 1.094$~AU, $\Delta_s = 0.232$~AU,
$\phi=63$\degr). The 8.0~\micron{} images saturated on the nucleus,
but the remaining short exposure (0.6~s) 3.6--5.8~\micron{} IRAC
images did not saturate and were used in our analysis
(Fig.~\ref{nov-encke-images}). Encke was again imaged with IRAC during
normal science operations \dataset[ADS/Sa.Spitzer\#6581760]{(PID~119,
AORKEY~6581760)} on 2004 June 29 at 22:05~UT ($r_h = 2.611$~AU,
$\Delta_s = 1.958$~AU, $\phi=22$) and with the Multiband Imaging
Photometer for \textit{Spitzer} (MIPS) at 24 and 70~\micron{}
\dataset[ADS/Sa.Spitzer\#6582272]{(PID~119, AORKEY~6582272)} on 2004
June 23 at 05:04~UT ($r_h = 2.556$~AU, $\Delta_s = 1.997$~AU,
$\phi=20$\degr; Fig.~\ref{jun-encke-images}).  Detailed information
regarding the performance and use of IRAC and MIPS is provided by
\citet{fazio04} and \citet{rieke04}, respectively.  The imaging
observations are summarized in Table~\ref{image-log}.  The 2003
November data were processed with IRAC pipeline S11.0.2 and the 2004
June data were processed with IRAC and MIPS pipelines S11.4.0.

The 2004 June comet Encke images were mosaicked in the rest frame of
the comet with the MOPEX software \citep{makovoz05} at the native IRAC
and MIPS scales (1.22\arcsec~pixel$^{-1}$ for IRAC,
2.5\arcsec~pixel$^{-1}$ for MIPS 24, and 5.1\arcsec~pixel$^{-1}$ for
MIPS 70) and are presented in Fig.~\ref{jun-encke-images}.  The
nucleus clearly dominates the 5.8, 8, and 24~\micron{} images based on
the tenuous coma emission and distinct diffraction rings.  Careful
inspection also reveals that the nucleus is detected at 3.6, 4.5, and
70~\micron.  Comet Encke's debris trail is seen as a diagonal line of
emission extending across the 8 and 24~\micron{} images
\citep{gehrz06}.  Also evident are two streaks of material that
comprise the remains of Encke's activity during this perihelion
passage.

The IRS peak-up images are presented in Fig.~\ref{peakups}.  All
images use the red peak-up array ($\lambda_{c} = 22$~\micron{}) except
the 2004 June 25 08:27~UT Encke peak-up, which uses the blue peak-up
array ($\lambda_{c} = 16$~\micron).  The images are not flux
calibrated but are useful in describing the comet morphologies in the
following sections.

\section{RESULTS AND MODELS}
\subsection{2P/Encke}
\subsubsection{The Nucleus and the NEATM}\label{encke-results}
The spectrum of the Encke nucleus must be estimated and subtracted
from the raw comet spectrum before extended source calibration, as
discussed in \S\ref{spectra-obs-text}.  The light from a comet nucleus
is a combination of reflected light and thermal emission at the
shortest wavelengths ($\lesssim5$~\micron) and solely comprised of
thermal emission at longer wavelengths ($\gtrsim5$~\micron).  We
approximate the reflected solar spectrum with a 5770~K black body.
The thermal emission for comet nuclei is typically modeled with
asteroid thermal models \citep{campins87, veeder87, fernandez00,
stansberry04}, which have consistently modeled the thermal emission
from those comets visited by spacecraft, for example 9P/Tempel~1
\citep{lisse05, harker05} and 81P/Wild~2 \citep{fernandez99}.
Originally, the standard thermal model (STM) was successful in
modeling the thermal emission from asteroids at small ($\le30$\degr),
phase angles \citep{lebofsky86}.  \citet{harris98} proposed the
near-Earth asteroid thermal model (NEATM) that extends the STM to
higher phase angles and introduces a variable infrared (IR) beaming
parameter, $\eta$, rather than holding it constant at 0.756; a value
which reproduces the diameters of asteroids Ceres and Pallas
\citep{lebofsky86}.  The IR beaming parameter in models either raises
or lowers the sub-solar temperature, $T_{ss}$, as
\begin{equation}
T_{ss} = \left[\frac{(1 - A) S}{\eta \epsilon \sigma}\right]^{1/4}\ 
\mbox{(K)},
\end{equation}
where $A$ is the bolometric bond albedo, $S$ is the incident solar
flux, $\epsilon$ is the infrared emissivity, and $\sigma$ is the
Stefan-Boltzmann constant.  In our treatment, a scaled 5770~K black
body spectrum was added to account for any reflected light at the
shortest wavelengths ($\lesssim 5$~\micron),
\begin{equation}
F_\lambda = \frac{\alpha \pi B_\lambda(5770~\mbox{K}) R^2 R_\sun^2}{4
\Delta^2 r_h^2}\ \mbox{(W~cm}^{-2}~\mbox{\micron}^{-1}\mbox{)},
\end{equation}
where $\alpha$ is freely variable, unit less scale parameter, $\pi
B_\lambda$ is the Planck function in W~cm$^{-2}$~\micron$^{-1}$, $R$
is the radius of the comet in km, $R_\sun$ is the radius of the sun in
km, $\Delta$ is the comet-observer distance in km, and $r_h$ is the
comet-sun distance in km.

The near diffraction limited \textit{Spitzer} images of comet Encke
can be used to estimate the size of the comet nucleus if the coma
contribution to the observed surface brightness is negligible or can
be removed.  In the two sets of IRAC images and the MIPS 70~\micron{}
image (Fig.~\ref{nov-encke-images} and \ref{jun-encke-images}), the
coma was faint or undetected and the nucleus was sufficiently measured
through aperture photometry.  Aperture corrections from the IRAC Data
Handbook were applied to the IRAC photometry to account for the chosen
aperture radii and background annuli [IRAC is calibrated with a
10~pixel radius aperture and a 10--20~pixel annulus, \citep{idh}].
Aperture corrections applied to the MIPS photometry were derived from
the 500~K MIPS point source response functions provided by the
SSC\footnotemark[3].  All apertures were 5--10~pixels in radius except
for the 2004 June IRAC 3.6 and 4.5~\micron{} apertures, which were set
to 2 pixels in radius to avoid neighboring stars.  The MIPS
24~\micron{} image was fitted with a point source derived from stars
in the field and the IRAC 8~\micron{} image was saturated by the
nucleus.  The measured fluxes are provided in Table~\ref{encke-nuke}.
The current estimate of the flux calibration errors is $\approx2$\%
for IRAC \citep{reach05} and $\approx10$\% and $\approx20$\% for MIPS
24 and 70 \citep{mdh}.  The errors in Table~\ref{encke-nuke} were
produced from the quadrature addition of the formal photometric errors
and the flux calibration errors.  \footnotetext[3]{Available at
http://ssc.spitzer.caltech.edu/mips/psffits/.}

The IRAC color corrections to the photometry are very sensitive to the
underlying spectral energy distribution (SED) when dealing with
black bodies in the 200--400~K range, e.g., the 3.6~\micron{} color
correction varies from 6--51\% for $T=400$--200~K \citep{idh}.
Additionally, the 3.6~\micron{} bandpass likely includes an equal
combination of thermal emission and reflected solar radiation from the
nucleus ($\approx50$\% from each) [e.g., the SED of comet C/1995~O1
(Hale-Bopp)---\citep{harker02}].  We fit the SED with an initial NEATM
model to approximate the true SED and calculate color corrections as
prescribed in the IRAC Data Handbook \citep{idh}.

The color corrected SEDs were modeled with a non-linear reduced
chi-squared ($\chi^2_\nu$) fitting routine using the NEATM to derive
the the effective radius of comet Encke's nucleus.  The 2004 June
epoch had the best spectral coverage, ranging from 3.6--70~\micron.
Three parameters affect the color-temperature of the nucleus,
$\epsilon$, $A$, and $\eta$.  For Encke's nucleus, we varied $\eta$ as
a free parameter, adopted a geometric albedo, $p_v = 0.047$, which is
related to the Bond albedo \citep{hanner81}, following the discussion
by \citet{fernandez00}, and set $\epsilon = 0.9$.  The best-fit
effective radius was $R = 2.34\pm0.14$~km ($\eta = 0.735\pm0.046$,
$\chi^2_\nu = 3.5$).  The entire list of fitted parameters is
presented in Table~\ref{encke-fit-table} and the model SED is
presented in Fig.~\ref{encke-fit-fig1}.

The 2003 November SED was limited in spectral coverage to
3.6--5.7~\micron{} and the number of photometry points (3) permitted
us to only fit two parameters at a time.  Therefore, we fixed the
value of $\eta$ to the best-fit value derived from the 2004 June epoch
and fitted $R$ and the scale parameter of the reflected light
($\alpha$).  The result was an effective nucleus radius of $1.72 \pm
0.10$~km with $\chi^2_\nu = 11.6$.  The error was derived by varying
$\eta$ over the 2004 June range of $\pm 0.046$.  The fit is presented
in Table~\ref{encke-fit-table} and the SED in
Fig.~\ref{encke-fit-fig2}.

The 2003 November and 2004 June best-fit effective nucleus radii are
different by 0.6~km.  If this is evidence for a non-spherical nucleus
then the axial ratio is at least $a/c = 1.4$.  This value is
consistent with the \citet{fernandez00} axial ratio of $a/c \ge 2.6$
derived from the visual light curve of the nucleus.  Our
\textit{Spitzer} observations were separated by seven months, however,
we can not reliably re-phase our data as the rotation period of Encke's
nucleus is not precisely known \citep{fernandez05, harmon05}.
Previous investigations of comet Encke's nucleus with thermal
observations provide effective radii consistent with our derived value
of $2.34\pm0.14$~km.  \citet{fernandez00} derive a value of
$R=2.4\pm0.3$~km, \citet{gehrz89} posit $R=2.5$--6.4~km, and
\citet{campins88} conclude $R<2.9$~km.

The 0.6~km effective radius disagreement in our \textit{Spitzer}
estimates can be minimized by varying $\eta$ in the 2003 November fits
and forcing the radius to be constant at 2.34~km.  This method results
in $\eta = 1.026 \pm 0.061$ and $\chi^2_\nu = 0.11$.  The new fit is
presented in Table~\ref{encke-fit-table} and in
Fig.~\ref{encke-fit-fig2}.  The low $\chi^2_\nu$ indicates an
under-constrained fit, but the fit appears improved as the scattered
light contribution in the model SED is similar to the 2004 June
best-fit (Fig.~\ref{encke-fit-fig1}).  Recent work on near-Earth
asteroids suggests $\eta$ may have a phase angle dependence
\citep{delbo03, wolters05}.  We discuss the radius of comet Encke and
the role of the IR beaming parameter further in \S\ref{eta-discuss}.

\subsubsection{The Coma and the Dust Thermal Model}\label{encke-coma}
To properly model and interpret the coma dust emission in comet Encke
requires a useful photometric calibration of the narrow slit spectra.
First, we must take into account the varying slit widths amongst the
IRS modules (ranging from 3.7\arcsec{} to 11.1\arcsec), normalizing to
an aperture size of radius 4.5\arcsec.  Our adopted aperture size is
roughly equivalent in area to the module with the smallest slit and
extraction aperture (SL), which is $3.7\arcsec{} \times 17.5$\arcsec{}
as described above (\S\ref{spectra-obs-text}).  Note that the aperture
normalization cannot account for the different regions sampled by the
different slit sizes and orientations (\S\ref{spectra-obs-text}).

Small pointing offsets will cause mis-calibrations of the spectral
flux, especially with the SL module, which has a slit width of
3.7\arcsec.  This effect is evident when we consider that the inner
peak of the \textit{Spitzer} point spread function (PSF) has a
full-width at half-maximum of about 2.4\arcsec{} at 10~\micron.  The
coma flux can be estimated from the MIPS 24~\micron{} image where we
have subtracted the nucleus.  The resultant flux can then be used to
photometrically calibrate all spectra through module-to-module
wavelength overlaps.  We measure the coma flux using two methods.
Aperture photometry (Method~1) on the residual coma of the point
source subtracted image yields a coma flux density of $3.36 \pm 0.20
\times 10^{-21}$~\wcm{} in a 4.5\arcsec{} radius (1.8 pixels) beam.
The error includes the photometric uncertainty of the nucleus
subtraction assuming it is distributed evenly over the PSF.
Alternatively (Method~2), we fit the azimuthally averaged aperture
flux density profile from 3--12~pixels in radius with a power-law,
$F(\rho) = C\rho^k$, where $C$ is a scale factor to account for the
total brightness, $\rho$ is the aperture radius in pixels, and $k$ is
the logarithmic slope.  Our fit yields the values $C =
1.46\pm0.03$~\wcm{} and $k = 1.509\pm0.009$ ($\chi^2_\nu = 0.8$).  The
best-fit slope is different than the nominal ``steady-state'' profile
\citep[$\rho^{1.0}$,][]{jewitt91} because of the highly structured
dust morphology of comet Encke at this epoch and these aperture radii
(see Fig.~\ref{jun-encke-images}).  Computing the flux enclosed in a
4.5\arcsec{} radius aperture with this surface brightness profile
yields a coma flux density of $3.55\pm0.18 \times 10^{-21}$~\wcm,
again including the PSF fit error.  The two methods are in good
agreement.  We elect to adopt Method~2 for scaling the IRS spectra as
it is less likely influenced by the pixel-to-pixel errors of the PSF
fitting at the nucleus removal step.  A cut across the PSF subtracted
image and the aperture photometry profile is presented in
Figs.~\ref{encke-psf1} and \ref{encke-psf2}.

The best-fit NEATM nucleus SED ($R = 2.34 \pm 0.14$~km) was subtracted
from the extracted comet Encke spectra and then scaled to match the
MIPS coma photometry.  First, we scaled the separate modules by area
to the fiducial 4.5\arcsec{} aperture.  Next, we scaled the spectra to
match the spectra overlap regions.  Finally, we integrated the
spectrum under the MIPS 24~\micron{} bandpass to determine the final
scale factors (Table~\ref{scales}).  The SL spectrum from the 2004
June 25 05:28~UT epoch was dominated by the errors after nucleus
subtraction (i.e., the coma was at low signal to noise), hence the
large scale factor of 1.53.  The final, combined 2004 June 25 05:28~UT
epoch spectrum is presented in Fig.~\ref{encke-spec}.

We repeat this procedure for the 2004 June 25 08:27~UT epoch,
high-resolution spectrum of Encke.  The NEATM nucleus predicts a flux
larger than observed in the SL spectrum data points by more than one
standard deviation (overall, $\chi^2_\nu = 22$).  An improperly
centered (within the slit) comet could cause such a photometric error,
or the nucleus may be presenting a larger cross section to the
telescope than the $R=2.34$~km model provides.  We cannot estimate the
coma flux in this module and therefore drop SL from the analysis.  The
SH and LH scale factors (Table~\ref{scales}) account for the
asymmetric dust morphology of comet Encke and the shape and
orientation of the slits.  The final spectrum is presented in
Fig.~\ref{encke-spec}.

The comet Encke spectra are modeled using a thermal grain model
developed by \citet{harker02, harker06} which self-consistently
calculates the temperature and thermal emission from a cometary dust
mixture using laboratory optical constants.  The minerals include
amorphous carbon, amorphous pyroxene, amorphous olivine, and
crystalline olivine.  The amorphous carbon is used to represent warm,
featureless continuum from deeply absorbing grains.  The amorphous
pyroxene and amorphous olivine grains are a 50-50 mixture of magnesium
and iron which have an increased temperature as compared to
pure-magnesium grains.  The increased temperature was required to
adequately model the silicate features in the thermal spectrum of
comet Hale-Bopp \citep{harker99a, harker02}.  The crystalline olivine
grains are magnesium-pure (forsterite).

The thermal model invokes a Hanner grain size distribution
\citep[HGSD;][]{hanner83} to calculate the emission from a population
of grains, $n(a)da$ with radii, $a$, varying from 0.1--100~\micron{}.
The grain population is described by a modified power law,
\begin{equation}
n(a) = \left(1 - \frac{a_0}{a}\right)^M \left(\frac{a_0}{a}\right)^N,
\end{equation}
where $a$ is the grain radius, $a_0$ is the minimum grain radius
(assumed to be 0.1~\micron), $N$ is the slope of the distribution at
large $a$, and $M$ is related to the radius of the peak of the grain
size distribution, $a_p$, by
\begin{equation}
a_p = a_0 \frac{M + N}{N}.
\end{equation}
This choice of grain size distribution is known to reproduce the
observed SEDs of many comets in the 3.5--20~\micron{} wavelength
range.  Model grains can be fractally porous, with a density described
by
\begin{equation}
\rho(a) = \rho_0\left(\frac{a}{a_0}\right)^{D-3},
\end{equation}
where $\rho_0$ is the bulk density and $D$ is the fractal dimension of
the dust (solid spheres have $D$ equal to 3, porous spheres have $D <
3$).  To keep coma fitting tractable, we chose discrete values of
$a_p$ (0.1~\micron{} steps) and $D$ [one of 3.0, 2.857, 2.727, 2.609,
and 2.5 \citep[see][]{harker02}], and all minerals in a given
observation are assumed to have the same grain size distribution.
After a best-fit model is derived, we attempt to estimate the error in
peak grain size by exploring the curvature of $\chi^2$ space with
respect to $a_p$.  For the Encke and HT50 best-fits below, the derived
$a_p$ errors are $\lesssim 0.01$~\micron.  Given the small formal
errors in $a_p$, our choice of 0.1~\micron{} steps ensures our derived
peak grain sizes are within 0.05~\micron{} from the best values.

The coma of comet Encke was fitted ($\chi^2_\nu = 0.5$--0.9) with a
porous ($D=2.857$), amorphous carbon mineralogy and a peak grain size
of 0.4~\micron{} ($N = 3.7$, $M = 11.1$).  The model spectra (solid
curve) are shown in Fig.~\ref{encke-spec}.  The agreement in best-fits
between the low-resolution and high-resolution data suggest our
reduction methods are consistent, regardless of the IRS module.  The
number of peak grains and upper-limits to olivine and pyroxene
minerals are presented in Table~\ref{encke-minerals}.  Our best-fit
model constrains the sub-micron silicate fraction to $<31$\% by mass.
The small peak grain size suggests that comet Encke has not exhausted
its reservoir of small particles, but the large particle slope
parameter, $N=3.7$, does not preclude the importance of large
particles.  For our derived PSD, the ratio of the total mass of
particles with $0.1~\micron{} \leq a \leq 1$~\micron{} to the total
mass of particles with $1~\micron{} \leq a \leq 10$~\micron{} is
0.076, i.e., the sub-micron particles are a minor component of the
total coma mass.  Discussion of comet Encke's mineralogy is presented
in \S\ref{mineral-discuss}.

\subsection{67P/Churyumov-Gerasimenko}\label{cg-results}
The peak up image of 67P (Fig.~\ref{peakups}) showed a point source
with a full-width at half-maximum of 3 pixels, or 18,000~km at the
distance of the comet.  Although the dust production of Jupiter-family
comets at 5~AU is generally assumed to be minimal or non-existent, a
point source does not indicate a bare nucleus.  A point source could
also be a combination of nucleus and 1) a coma from recent (hours to
weeks) activity, 2) a coma of slowly moving micrometer sized or larger
particles ejected this perihelion passage, or 3) very large particles
($\gtrsim 100~\micron$) entrained in 67P's debris trail.  Assuming
only a bare nucleus, we fitted the spectrum with the NEATM ($\eta =
0.756$, $p_v = 0.04$, and $\epsilon = 0.9$).  The fit yielded a
nucleus radius of $3.17\pm0.06$~km ($\chi_\nu^2 = 1.3$) and is
presented along with the spectrum in Fig.~\ref{cg-nuke}.

Our derived best-fit effective radius for this comet is inconsistent
with recent estimates of 67P's nucleus size \citep{lamy04}.  For
example, \citet{lamy03} derive a value of $1.98\pm0.02$~km using the
\textit{Hubble Space Telescope}, and \citet{kelley05} estimate a value
of $1.91\pm0.09$~km using \textit{Spitzer}/MIPS.  It is clear our
\textit{Spitzer} spectrum of comet 67P is not that of a bare nucleus.
Likely, some amount of dust still enshrouds the nucleus at 5~AU.  The
nucleus subtracted spectrum is not of high enough quality or spectral
range for a detailed coma fit.  When the \citet{kelley05} NEATM fit,
calculated for the epoch of the spectrum, is subtracted, the resulting
spectrum constrains the dust contribution.  The observed coma flux is
$2.01\pm0.10 \times 10^{-21}$~\wcm{} at 27.9~\micron{} (weighted
average from 21--35~\micron).

One component of 67P's dust emission at this large heliocentric
distance is the comet's debris trail.  To determine the fraction of
the flux originating from the trail, we use the trail parameters
determined by \citet{sykes92} from \textit{IRAS} measurements.  In the
\textit{Spitzer} observations, the trail contributes $5.1 \pm 0.6
\times 10^{-23}$~\wcm{}, or $2.5\pm0.3$\%, to the IRS spectrum.  We
can further refine this estimate by taking into account the effect of
the orbital motion at 5~AU in an eccentric orbit.  Here, particles in
similar orbits will move slower, and hence be closer together, than at
$r_h = 2.3$~AU (the epoch of the \textit{IRAS} observations).  To
estimate the optical depth enhancement caused by this effect, we
differentiate the true anomaly with respect to heliocentric distance
\begin{equation}
\frac{df}{dr} = \frac{-(1/e + \cos{f})}{r \sin^2{f}},
\end{equation}
where $f$ is the true anomaly at the time of observation ($f =
96.9$\degr{} at the IRAS epoch, $f = 156.0$\degr{} at the
\textit{Spitzer} epoch) and $e = 0.632$ is the eccentricity of the
orbit of comet 67P.  The ratio of optical depths from the
\textit{IRAS} epoch to the \textit{Spitzer} epoch is 1.25, increasing
the trail contribution to 3\%.  The majority of the dust emission
detected by \textit{Spitzer} likely arises from 1) recently ejected
dust (age of order hours to weeks), 2) large, slowly moving particles
from the 2003 perihelion passage [evidenced by the shallow
post-perihelion $r_h$ dependence of dust from optical observations
\citep{schleicher06}], or 3) some combination of the two.

The SED of 67P increases in flux density from 14--20~\micron{}, then
flattens (in slope) beyond 20~\micron{} (Fig.~\ref{cg-nuke}).  In
principle, the color-temperature of the coma can constrain the mineral
composition of the emitting dust.  We attempted to account for the
spectral shape by fitting the spectrum with our thermal model,
constrained to one mineral (amorphous olivine or amorphous carbon) and
constrained to specific peak grain sizes ($a_p = 1$, 5, 15, or
30~\micron).  For similarly sized particles, amorphous carbon dust has
a higher color-temperature than amorphous olivine, and thus produces a
poorer fit to the data.  Also, smaller sized particles are warmer and
produce poorer fits.  The best-fit model consisted of large ($a_p
\approx 15$~\micron) amorphous olivine grains.  For models of
amorphous olivine dust, $\chi^2_\nu$ ranged from 0.57 ($a_p =
15$~\micron{}, maximum grain size of 30~\micron) to 0.73 ($a_p =
1$~\micron{}, maximum grain size of 100~\micron).  The increased
maximum grain size in the $a_p = 1$~\micron{} model balances the
warmer, $\approx1$~\micron{} particles so that the resultant model
spectrum fits the observed color-temperature.  Overall, the low
signal-to-noise ratio of the SED and the small range of grain
temperatures for 1--100~\micron{} grains at 5.0~AU
($T\approx100$--150~K) prohibits a detailed discussion of 67P's dust
coma mineralogy.

\subsection{C/2001 HT50 (LINEAR-NEAT)}\label{ht50-results}
\begin{figure}
\plotone{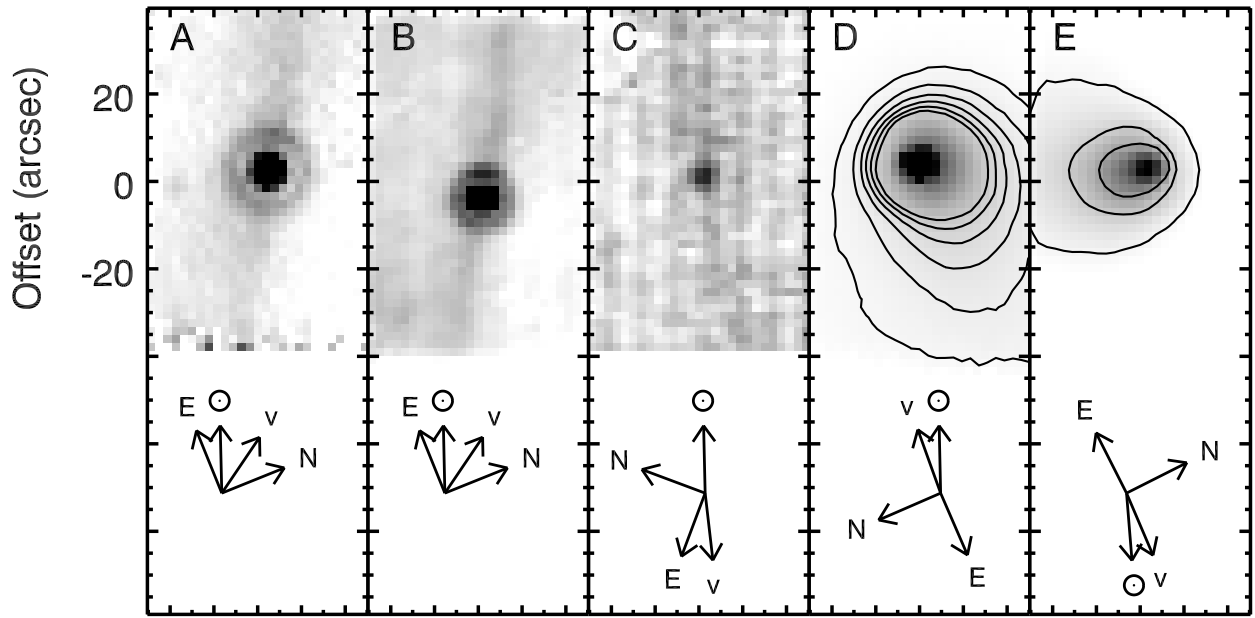}
\caption{IRS peak up images and orientations for all comets.  All
images are co-added standard S12 pipeline BCDs using the red peak-up
filter, unless noted: A) comet 2P/Encke, B) comet 2P/Encke, blue
peak-up filter, C) comet 67P/Churyumov-Gerasimenko, SL flat-field
applied, D) comet C/2001~HT50 (LINEAR-NEAT) at 3.2~AU with contours
spaced every +2000~DN starting with 10000~DN and E) comet C/2001~HT50
(LINEAR-NEAT) at 4.6~AU with contours spaced every +2000~DN starting
with 8500~DN.  The comet 2P/Encke peak-ups also show the ``spike''
features evident in the MIPS image of Fig.~\ref{jun-encke-images}.}
\label{peakups}
\end{figure}

\begin{figure}
\plotone{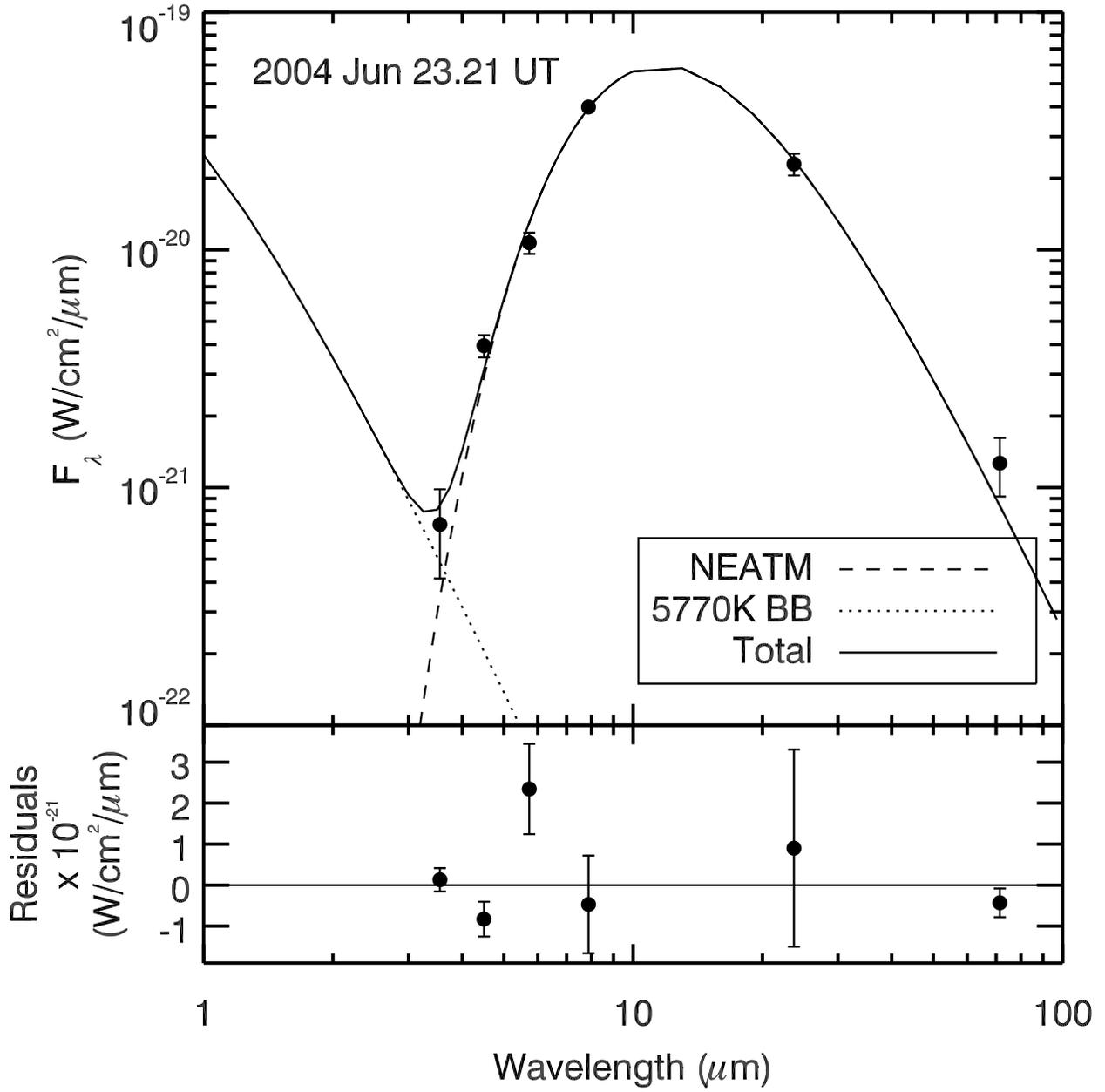}
\caption{NEATM fit to comet 2P/Encke's nucleus and residuals for the
2004 June epoch.}
\label{encke-fit-fig1}
\end{figure}

\begin{figure}
\epsscale{0.55}
\plotone{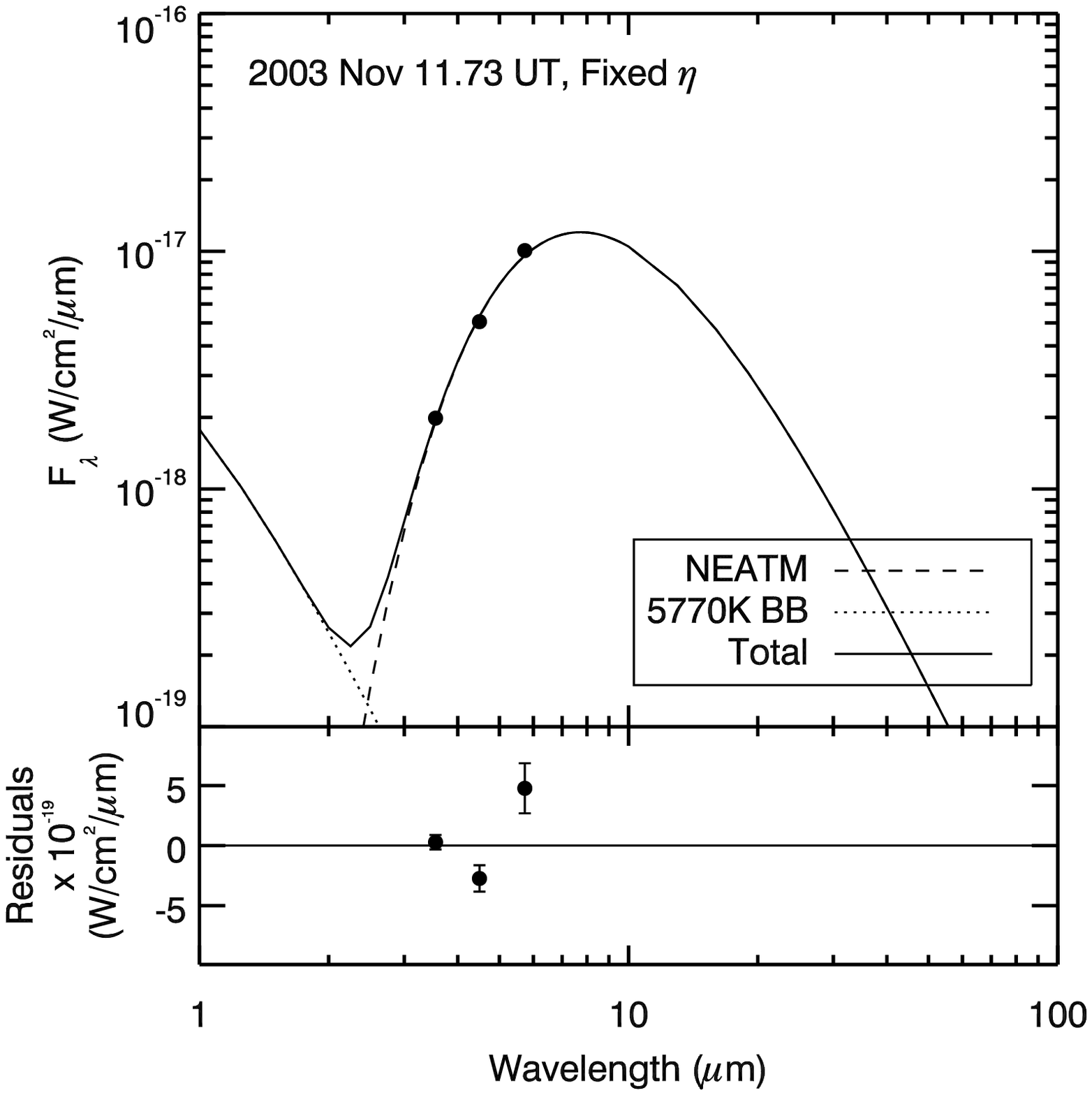}\\
\plotone{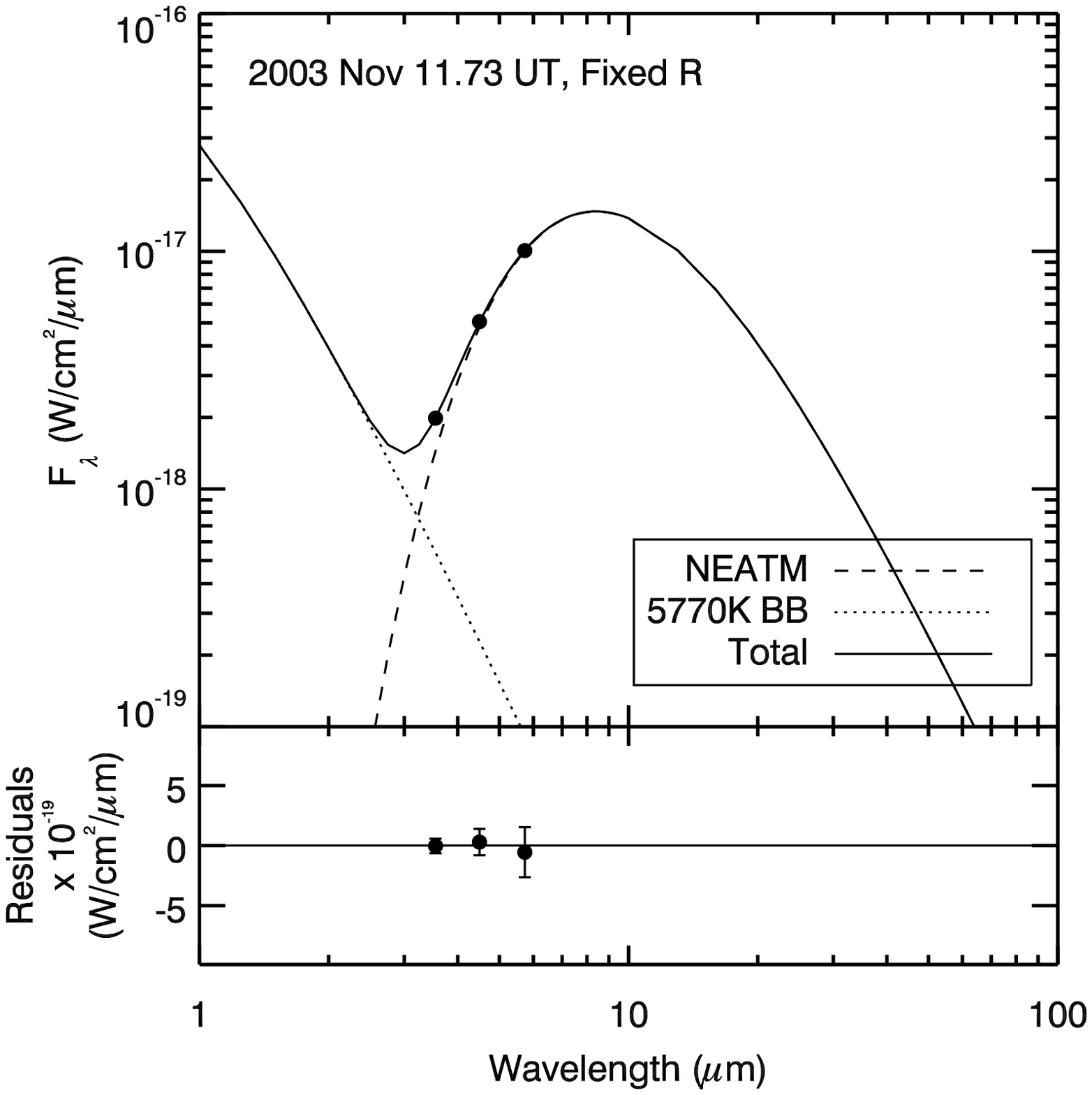}
\epsscale{1}
\caption{NEATM fits to comet 2P/Encke's nucleus for the 2003 November
epoch and their residuals.  The fits correspond to the fixed $\eta$
and fixed $R$ fits in Table~\ref{encke-fit-table}.}
\label{encke-fit-fig2}
\end{figure}

\begin{figure}
\plotone{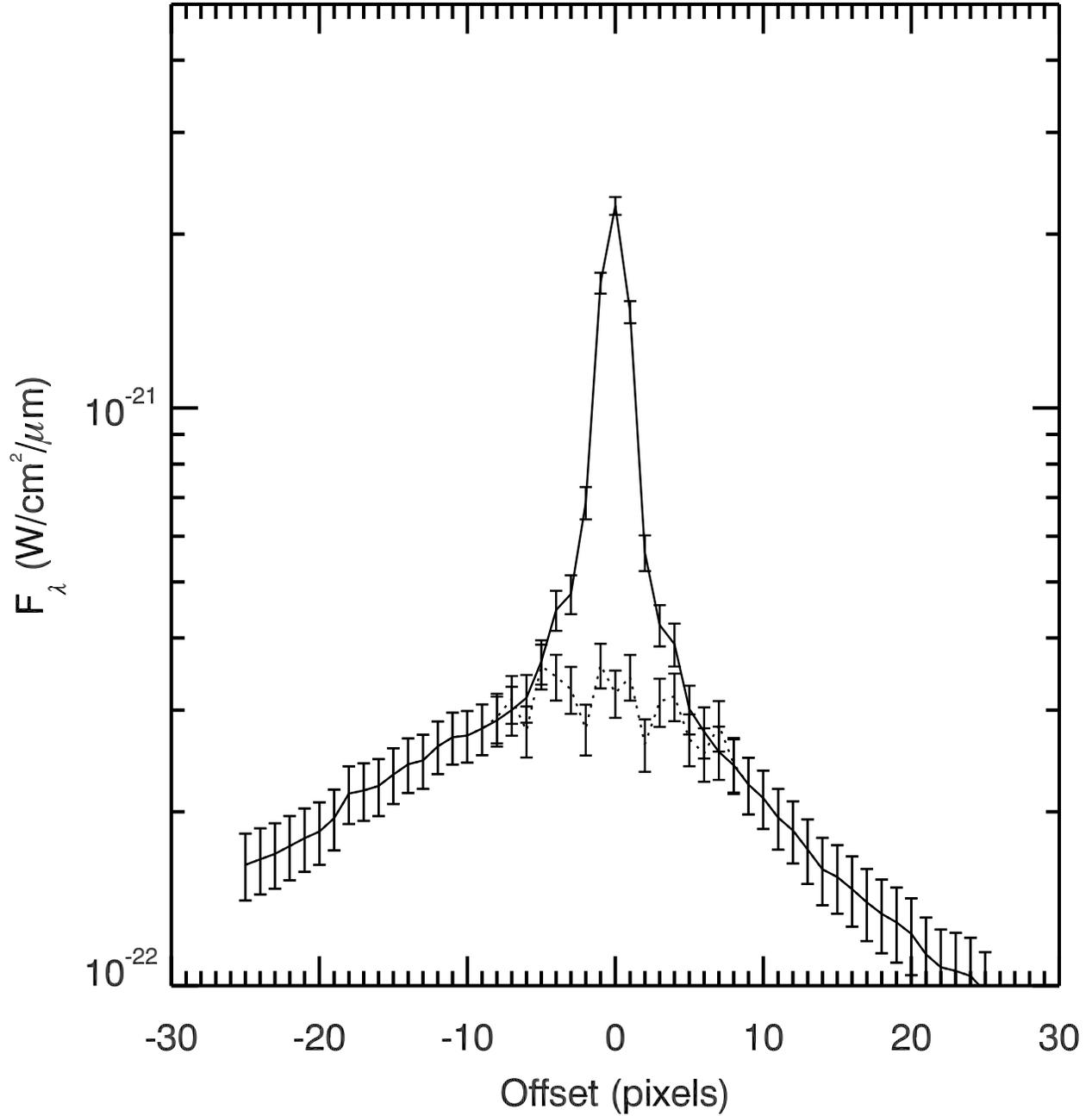}
\caption{Cuts along the spike features and the nucleus in the MIPS
image of comet 2P/Encke (position angle $\approx81$\degr; see
Fig.~\ref{jun-encke-images}).  The solid-line is extracted from the
original image, the dotted-line is extracted from the point source
subtracted image.}
\label{encke-psf1}
\end{figure}

\begin{figure}
\epsscale{0.95}
\plotone{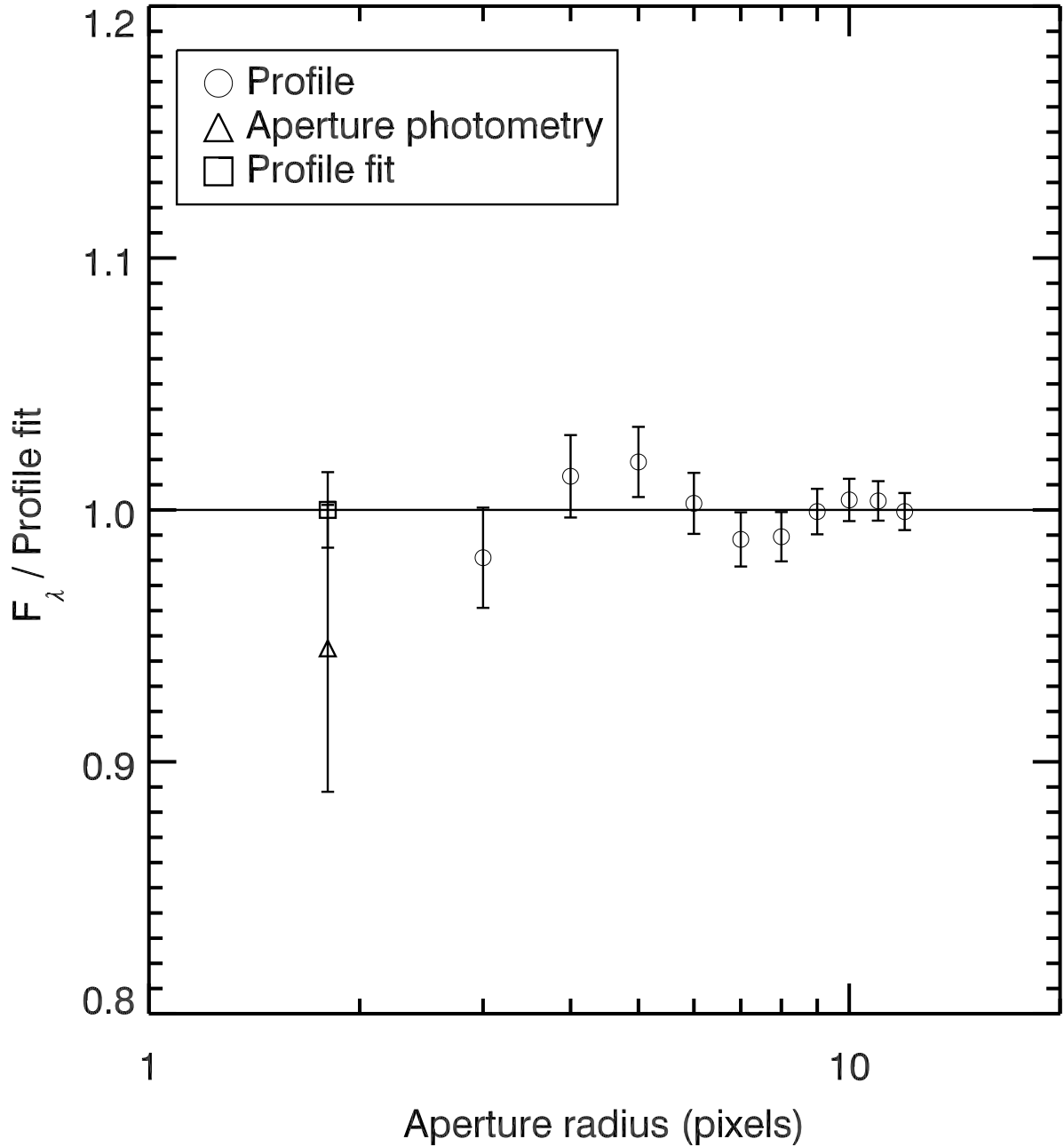}
\epsscale{1}
\caption{Azimuthally averaged aperture profile centered on the nucleus
of comet 2P/Encke and normalized to the profile fit.  The data points
at 1.8~pixels (4.5\arcsec) represent the coma fluxes in that aperture
as determined by aperture photometry and profile fitting.  The error
on the profile fit at 1.8~pixels ($\square$) does not yet include the
nucleus subtraction error.}
\label{encke-psf2}
\end{figure}

\begin{figure}
\epsscale{0.45}
\plotone{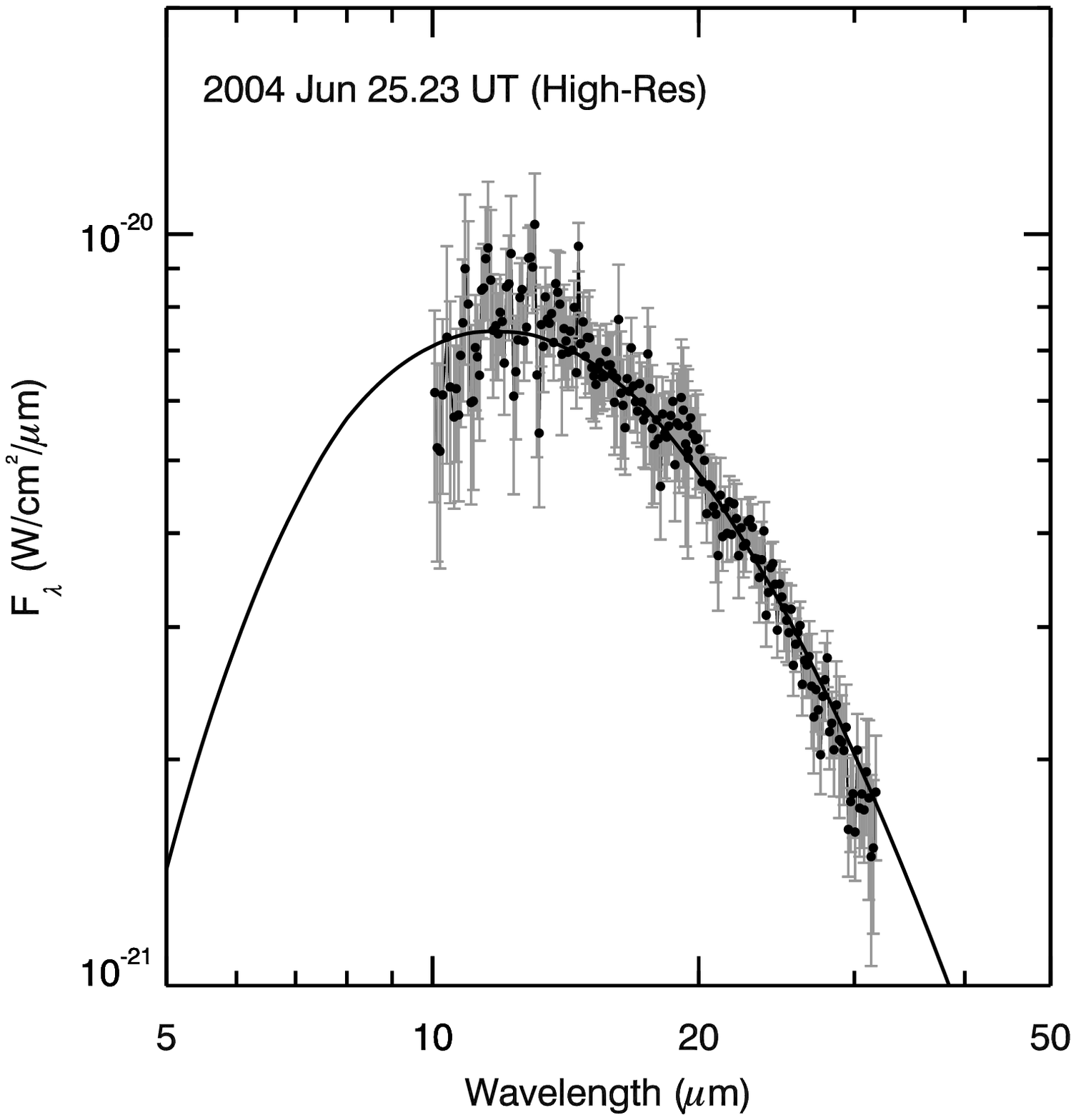}\\
\plotone{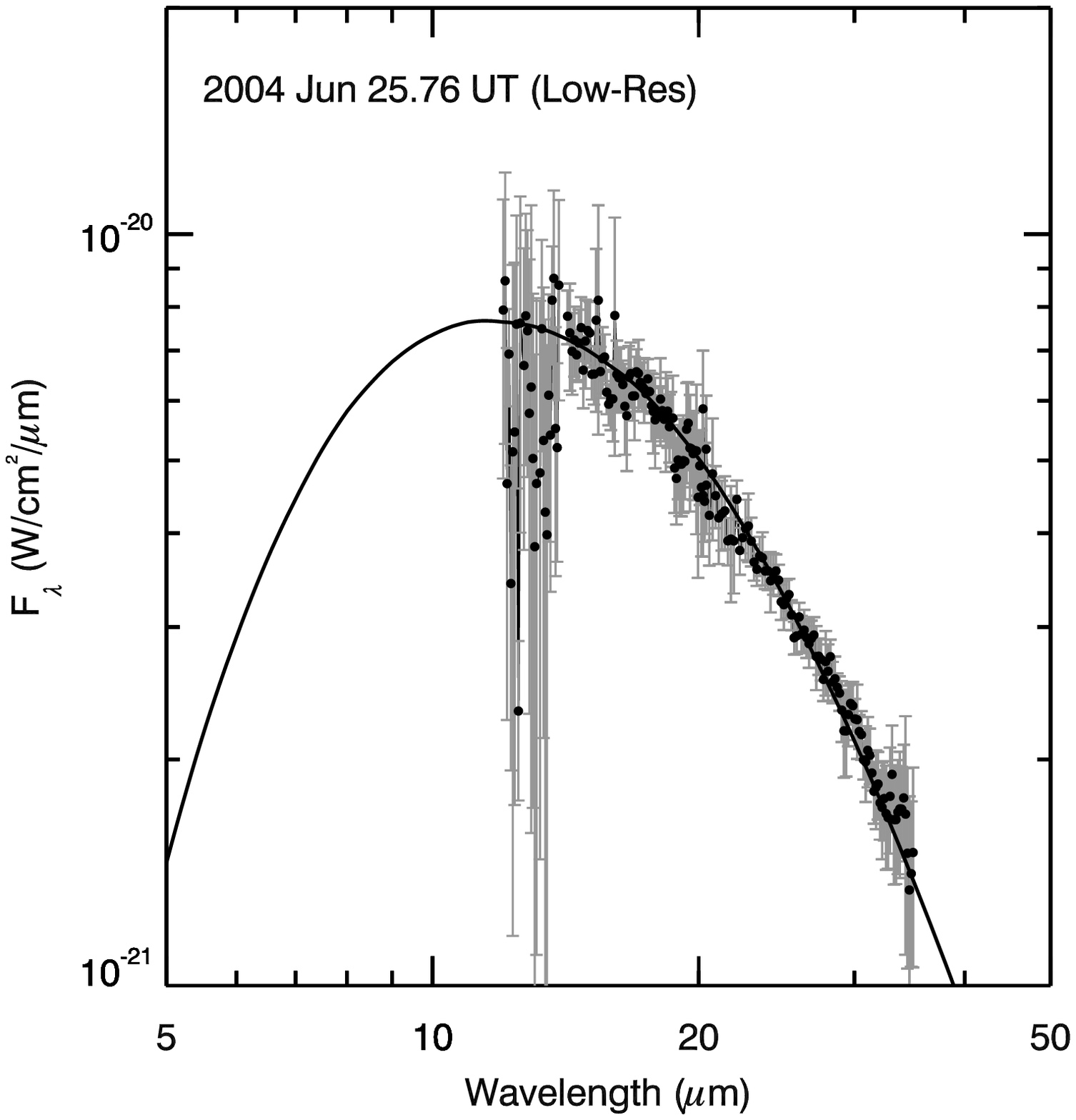}
\epsscale{1}
\caption{High-resolution and low-resolution IRS spectra of comet
2P/Encke.  The high-resolution spectrum was degraded from
$R\approx600$ to $R\approx85$ by a 7-point statistically weighted
average.  Both spectra were scaled to match the 24~\micron{} coma
photometry for a 4.5\arcsec{} radius aperture.  The best-fit thermal
emission models are shown as a solid-line ($N=3.7$, $M=11.1$,
$a_p=0.4$~\micron).}
\label{encke-spec}
\end{figure}

\begin{figure}
\plotone{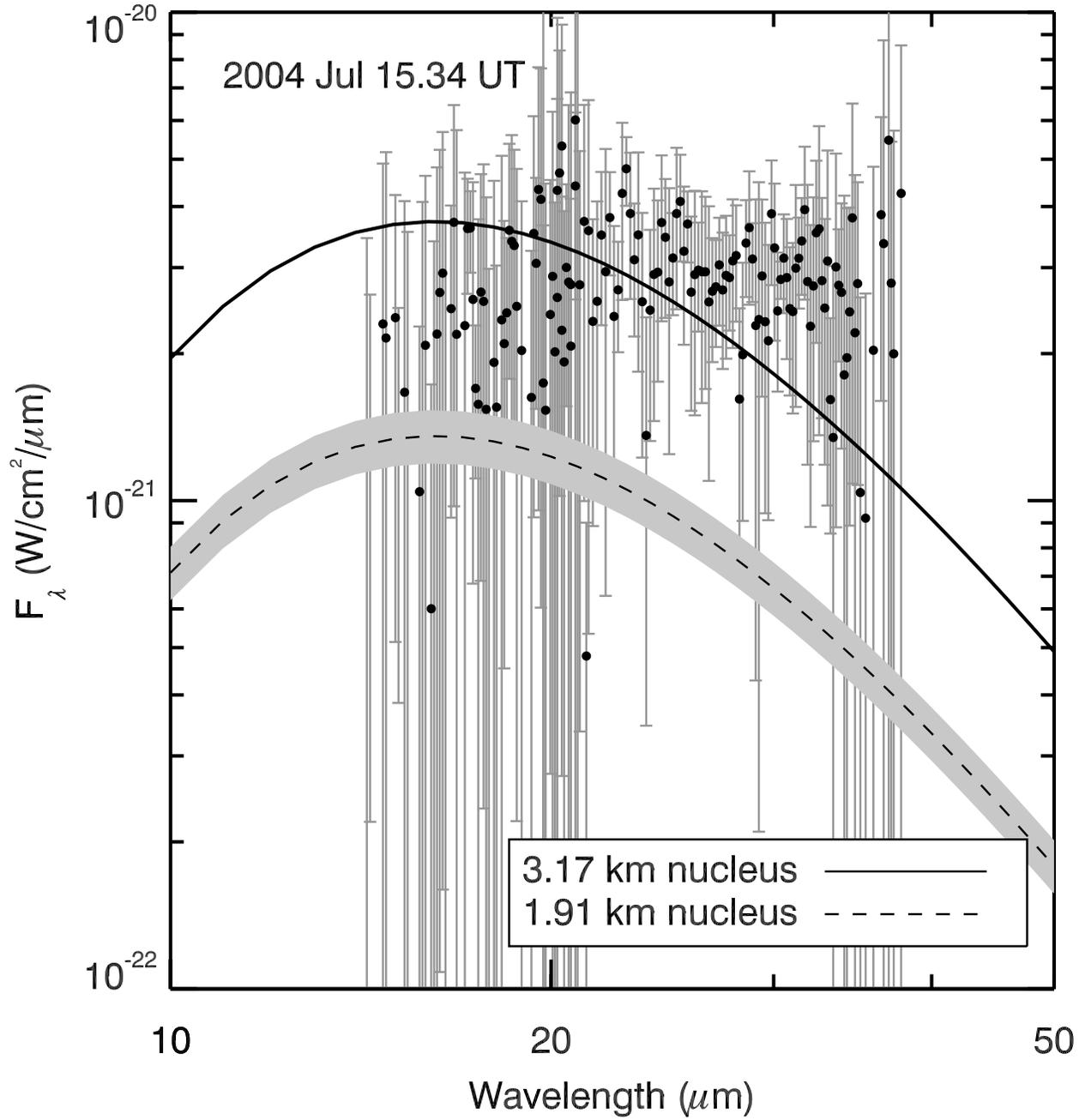}
\caption{IRS spectrum of comet 67P/Churyumov-Gerasimenko and model
nuclei.  Only data points with a signal to noise greater than 0.5 are
shown.  The shaded area indicates the one standard deviation error on
the nucleus size from \citet{kelley05}, $R = 1.91 \pm 0.09$~km.}
\label{cg-nuke}
\end{figure}

Inspection of the comet HT50 peak-up images (Fig.~\ref{peakups})
reveals a bright coma at both epochs ($r_h = 3.2$ and 4.6~AU).  We
inspected the peak-up images for the signature of a point source.  The
peak-up image at $r_h = 3.2$~AU is saturated over the center five
pixels and does not yield any information on the nucleus.  The profile
of the peak-up image at $r_h = 4.6$~AU does not exhibit a point source
within the coma profile.  We therefore proceed with the assumption
that the nucleus contribution at each epoch is negligible and do not
subtract a model nucleus from the HT50 spectra.

\begin{figure}
\epsscale{0.5}
\plotone{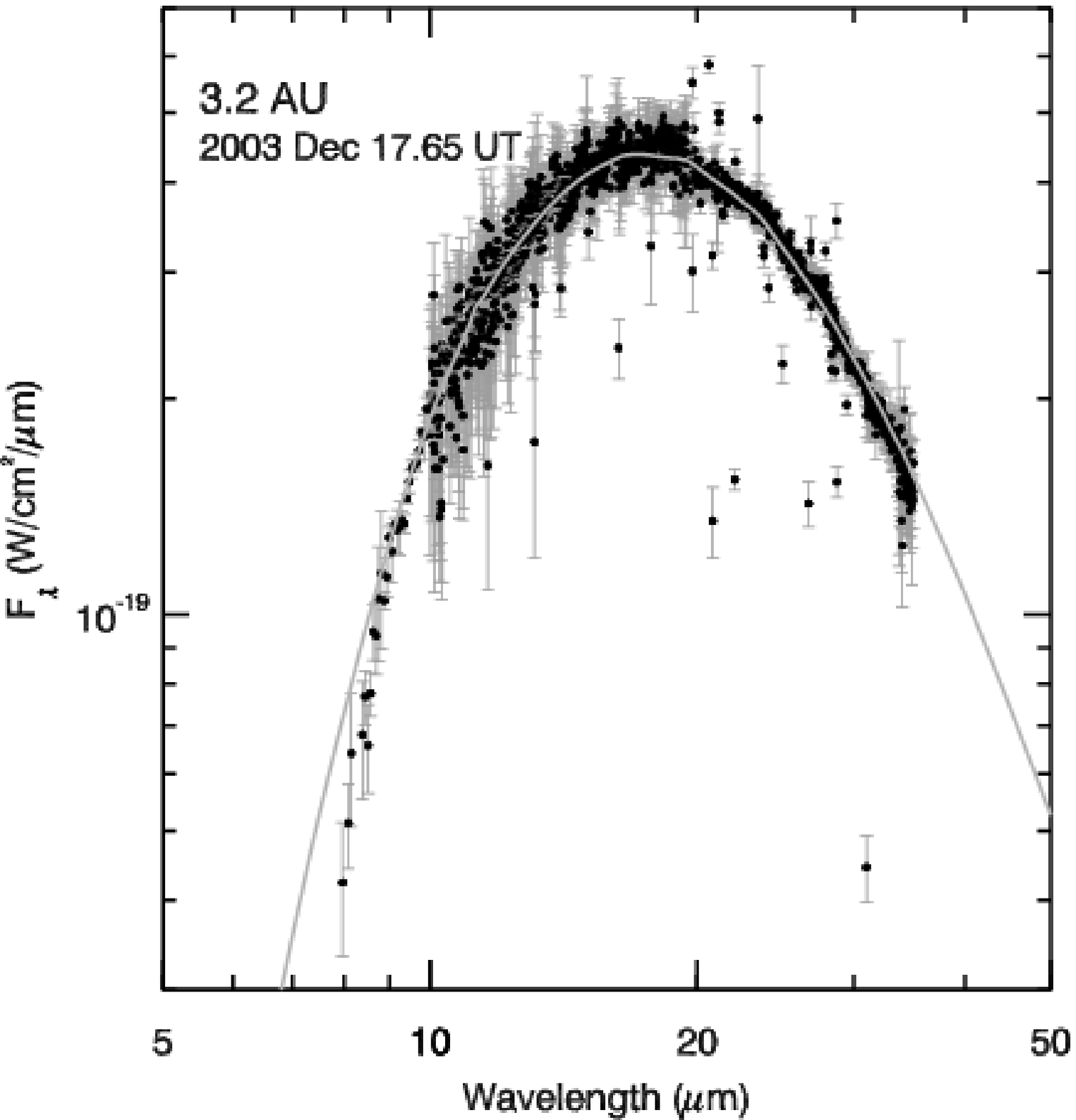}\\
\plotone{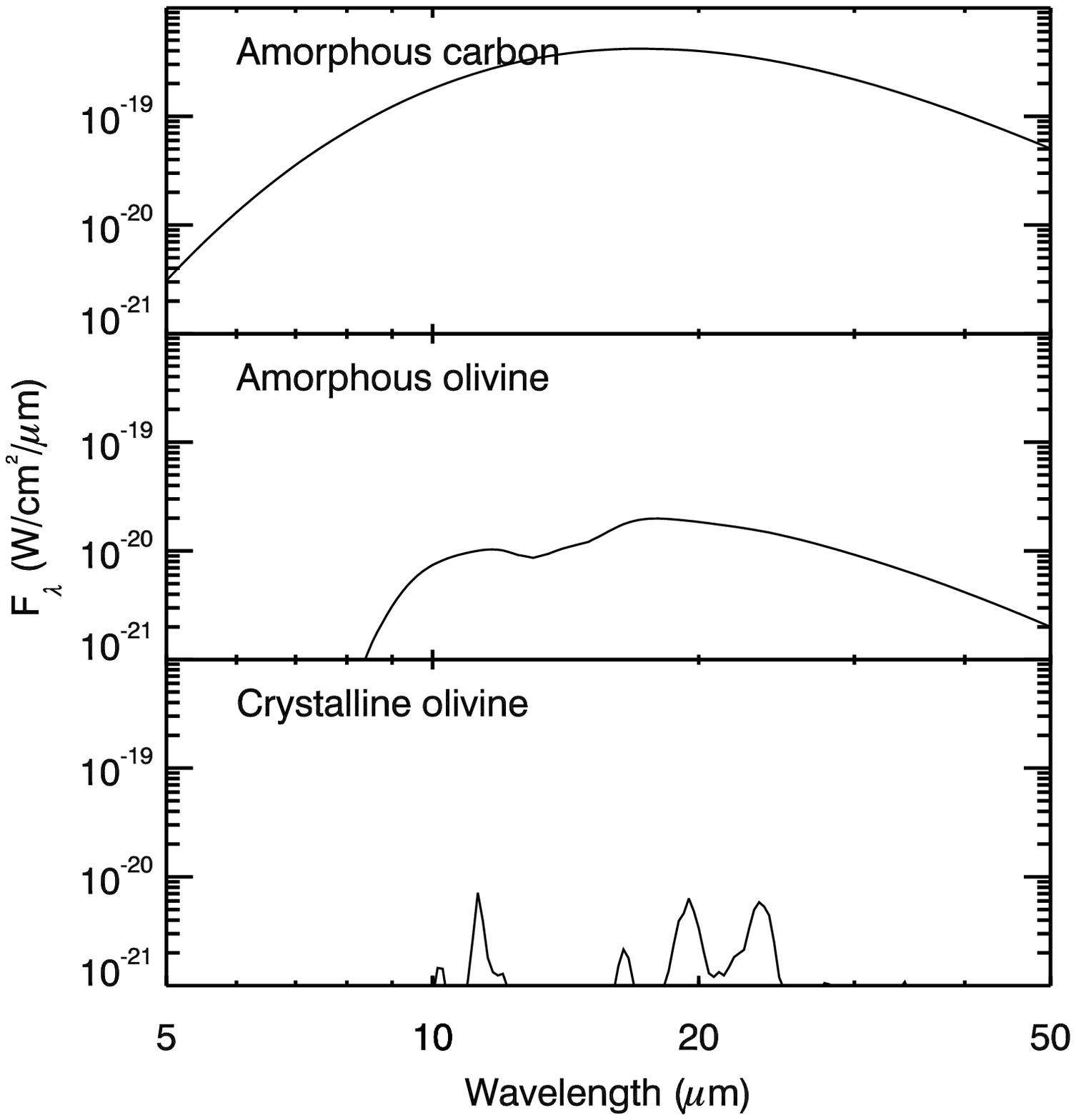}
\epsscale{1}
\caption{IRS spectrum of comet C/2001~HT50 (LINEAR-NEAT) at $r =
3.2$~AU and best-fit model of the dust thermal emission ($N=4.2$,
$M=46.2$, $a_p=1.2$~\micron).  Also shown are the decomposed spectra
for each of the significant minerals in the best-fit model.}
\label{ht50-3au}
\end{figure}

\begin{figure}
\epsscale{0.5}
\plotone{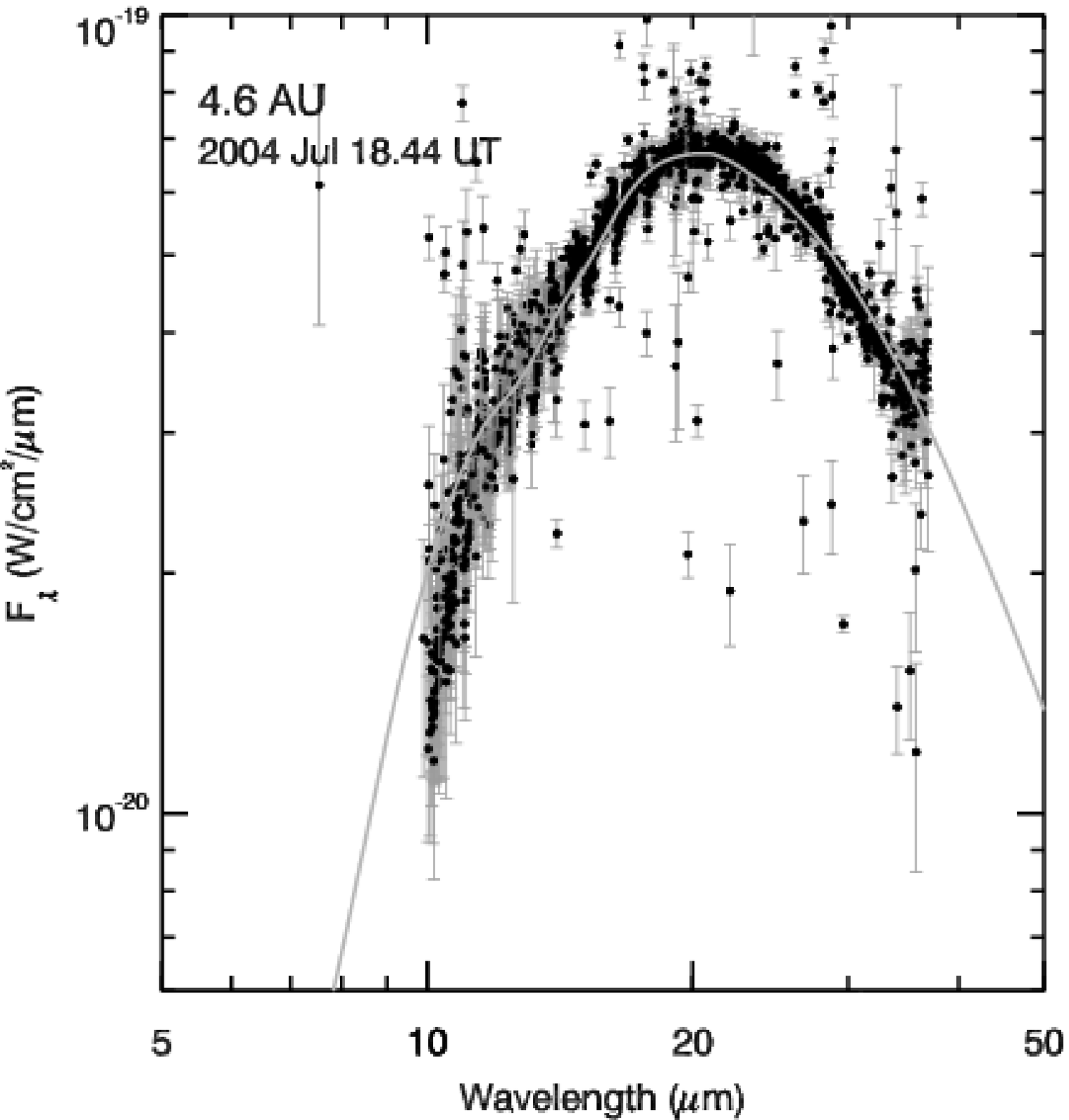}\\
\plotone{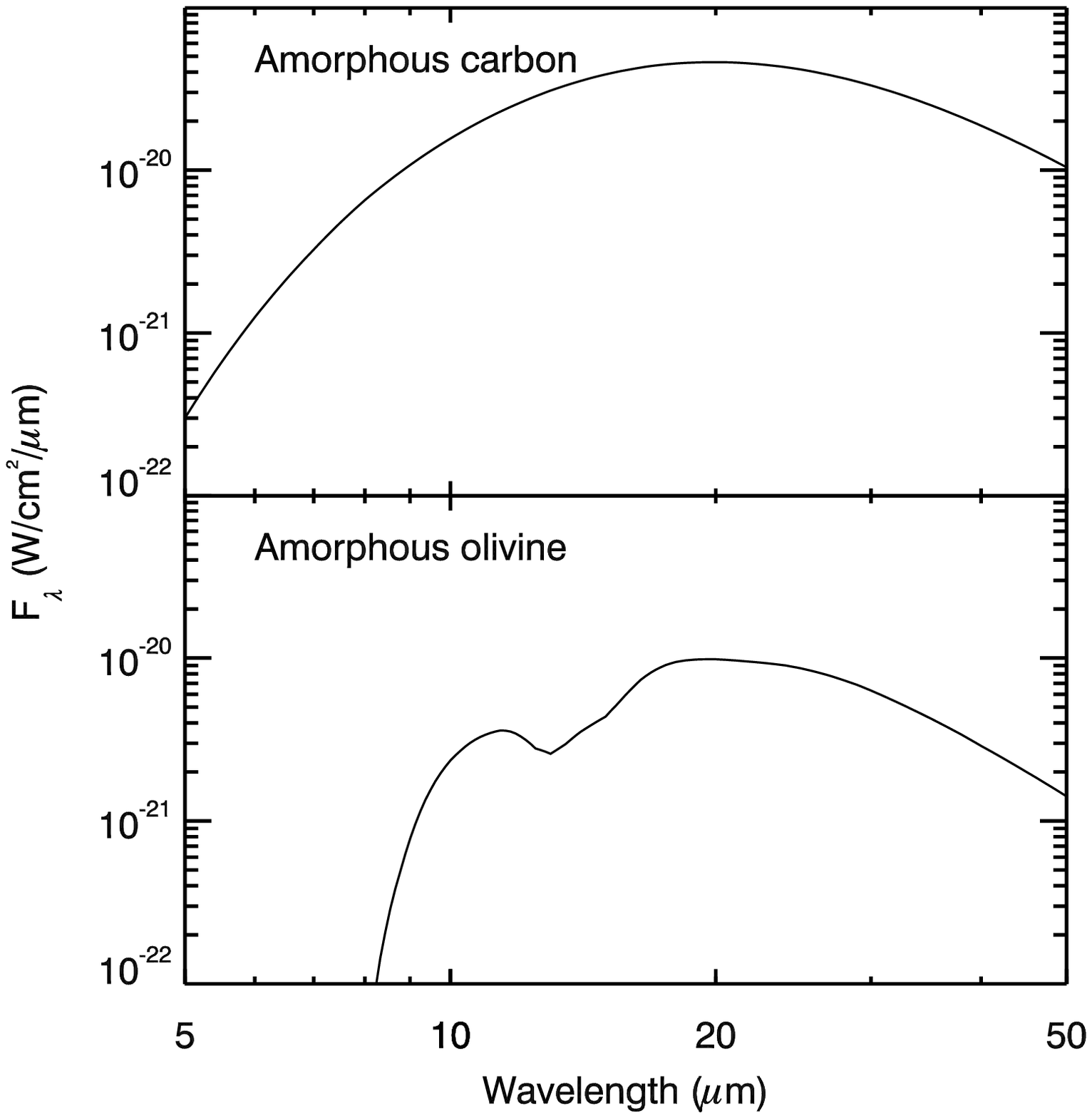}
\epsscale{1}
\caption{IRS spectrum of comet C/2001~HT50 (LINEAR-NEAT) at $r =
4.6$~AU and best-fit model of the dust thermal emission ($N=3.7$,
$M=29.6$, $a_p=1.2$~\micron).  Also shown are the decomposed spectra
for each of the significant minerals in the best-fit model.}
\label{ht50-4au}
\end{figure}

All IRS spectra were scaled to the nominal 4.5\arcsec{} aperture to
mitigate module photometric mis-matches.  Similar to the comet Encke
spectra, the extracted HT50 spectra must be further scaled to account
for varying slit orientations, coma asymmetries (Fig.~\ref{peakups}),
and the overall coma profile (Table~\ref{scales}).  The spectra and
our best-fit thermal models are presented in Figs.~\ref{ht50-3au} and
\ref{ht50-4au}, and in Table~\ref{ht50-minerals}.

The best-fit thermal models have a large peak grain size of
1.2~\micron{} and a mixed amorphous carbon and silicate mineralogy
($N=4.2$, $M=46.2$, $\chi^2_\nu=4.2$ at 3.2~AU and $N=3.7$, $M=29.6$,
$\chi^2_\nu=23.8$ at 4.6~AU).  The total number of amorphous carbon
grains from 0.1--10~\micron{} varies from $\approx5.8\times10^{19}$
($r_h = 3.2$~AU) to $\approx4.6\times10^{19}$ ($r_h = 4.6$~AU).  A
number of factors contribute to the number of observed grains,
including the changing grain size distribution, the aperture size at
the distance of the comet, the shape of the coma, and the dependence
of coma activity on heliocentric distance (including phenomena such as
jet activity and gas/dust outbursts).  We cannot account for the
heliocentric dependence of the dust production without a more rigorous
temporal sampling of the coma at these epochs.

\begin{figure}
\plotone{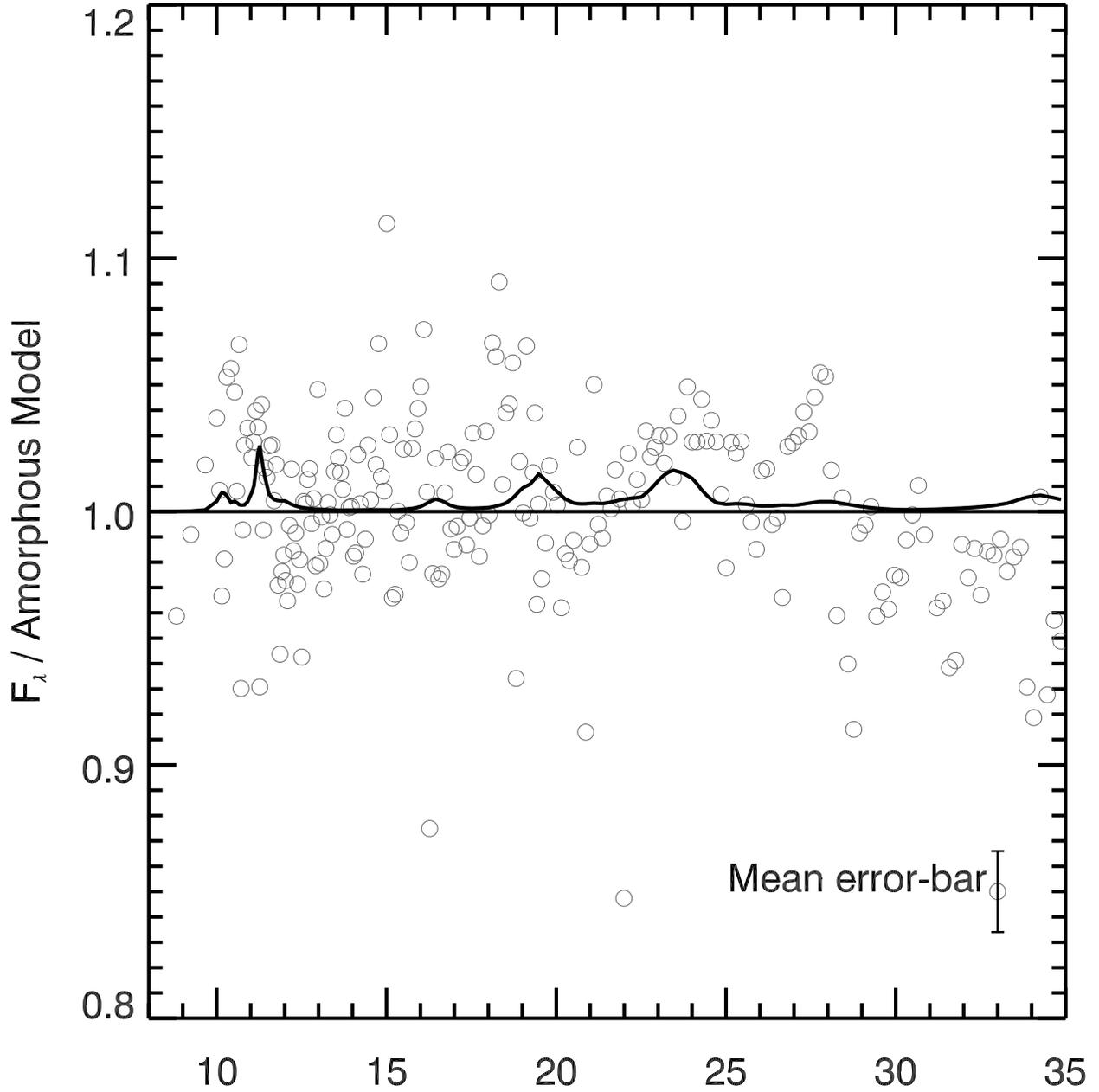}
\caption{Spectrum of comet C/2001~HT50 (LINEAR-NEAT) at $r = 3.2$~AU
binned by a seven-point statistically weighted, moving average and
normalized by the sum of our amorphous mineral models (open circles);
the solid line is the best-fit emission component from crystalline
olivine dust.  The presence of crystalline olivine as determined by
the model appears to be driven by the shape of the spectrum at
23.5~\micron.}
\label{ht50-cryst}
\end{figure}

Our best-fit model indicates a weak detection of crystalline olivine
in HT50 at 3.2~AU with a signal-to-noise ratio of 4.
Figure~\ref{ht50-cryst} presents a closer analysis of the crystalline
fit.  Here, the spectrum was smoothed by a seven-point statistically
weighted, moving average to increase the signal-to-noise of the
spectrum.  The spectrum was then normalized by the best-fit model,
excluding the crystalline component.  Excess emission (above unity in
the normalized spectrum) is potentially due to crystalline olivine
dust emission (represented in the figure by the solid line).  The
chi-squared fitting of the 1-sigma correlated errors indicates that a
better fit ($\Delta\chi^2_\nu = 0.04$) is obtained with crystalline
olivine present rather than absent (to 4-$\sigma$ or $>99\%$
confidence).  Inspection of Fig.~\ref{ht50-cryst} suggests the
detection of crystals is driven by the shape of the spectrum at
22--24~\micron.  Altogether, the \textit{Spitzer} observations suggest
the presence of crystalline olivine in comet HT50.  The mineralogy of
comet HT50 is discussed and compared to other Oort cloud comets in
\S\ref{mineral-discuss}.

\section{DISCUSSION}
\subsection{The IR Beaming Parameter}\label{eta-discuss}
Recent observations of near-Earth objects (NEO) suggest a possible
dependence of $\eta$ on phase angle \citep{delbo03, wolters05}.
\citeauthor{delbo03} derived an empirical $\eta$-phase angle
correlation from observations of NEOs at phase angles $\phi \approx
5$--60\degr.  \citeauthor{wolters05} added more data to the discussion
and derived the trend $\eta = 0.69 + 0.012 \phi$.  However, extensive
data to assess whether an $\eta$-phase angle correlation exists for
\textit{single} objects does not exist.  Only one asteroid (2002~NY40)
has been observed in detail at disparate phase angles and no
$\eta$-phase dependence was found \citep{muller04}.  Comet Encke is a
near-Earth object, and observations of its nucleus can be used to
assess the validity of the NEATM and the $\eta$-phase angle
correlation.

To examine the potential correlation of the IR beaming parameter with
phase angle, we will assume the nucleus has a radius equal to the
radius derived from our 2004 June SED, $R = 2.34 \pm 0.14$~km.  The IR
beaming parameter is also best constrained by the 2004 June SED to a
value of $0.735 \pm 0.046$.  With these parameters, the NEATM predicts
a flux of 3.9~Jy for the \citep{fernandez00} observation on 1997 July
19 at a phase angle of 44\degr.  However, \citeauthor{fernandez00}
observed a flux density of $2.74 \pm 0.24$~Jy.  The 2004 June derived
radius and IR beaming parameter do not account for the observed flux
in 1997 July, just as they did not account for the 2003 November SED
(\S\ref{encke-results}).  The \citeauthor{fernandez00} 1997 July 8.5,
10.7, and 11.6~\micron{} observations were fitted with the 2004 June
radius and we derived a best-fit $\eta$ value of $1.02 \pm 0.11$.  The
2003 November and 1997 July data sets are limited, but suggest a
larger $\eta$ value is required for higher phase angles.

An $\eta$-phase angle dependence can be a direct result of certain
physical properties of the nucleus surface.  There are two possible
mechanisms that may produce the a change of $\eta$ with phase angle
\citep[see][]{delbo03, wolters05}.  Either, the surface roughness of
the nucleus is exacerbated at high phase angles (i.e., the observer
views an increasing amount of shadowing), requiring a lower
temperature to model the nucleus; or alternatively, the night
hemisphere of the nucleus does not completely cool to the background
temperature and contributes appreciable flux at high phase angles,
requiring a lower temperature to model the nucleus.  We consider these
possibilities in order.

In the former case, surface roughness causes severe shadowing across
the diurnal hemisphere.  At high phase angles, an observer could be
viewing the shadowed side of a scarp or other surface topology as
observed on the surfaces of comets 1P/Halley, Tempel~1, 19P/Borrelly,
and 81P/Wild~2 \citep{keller86, ahearn05, soderblom02, brownlee04}.
This geometry presents an overall cooler surface when compared to a
smooth sphere at the same phase angle \citep[see][Fig.~5]{delbo02}.  This
requires a lower temperature in the NEATM and, therefore, a higher IR
beaming parameter.  At low phase angles the contrary is true and an
observer may be viewing more surfaces normal to the sun direction,
thus requiring a hotter surface and lower IR beaming parameter in the
NEATM, consistent with the \citet{wolters05} trend.  The temperature
map of the Tempel~1 nucleus by \citet{ahearn05} clearly shows how
local surface temperature relies on the topology of the nucleus.
Surface roughness on smaller scales (e.g., in the regolith of small
bodies) is also important in thermal observations \citep{shkuratov00,
lagerros98} and necessary to explain polarization measurements of
atmophereless bodies \citep{petrova01}.

If surface elements on the night hemisphere contribute significant
flux to an observer, they can lower the observed color-temperature.
In the formalism of the NEATM, all radiation is assumed to arise from
the sun-lit hemisphere and therefore a lower observed
color-temperature will raise the derived $\eta$ value.  The
temperature of the night hemisphere will depend upon the angular
rotation rate of the nucleus and the ability of the surface to hold
heat, i.e., its thermal inertia.  Following \citet{spencer89} and
\citet{fernandez00} we can test whether or not the Encke nucleus could
be considered a fast or slow rotator.  A slow rotator model uses the
same assumptions as the STM and NEATM, i.e., the sub-solar point is
the hottest point and the night side emits no light.  A fast rotator
model assumes the cooling time scale is longer than the rotation period
and, therefore, the object will be isothermal with respect to
latitude.  The unit less parameter $\Theta$ is used to determine the
applicability of the two models,
\begin{equation}
\Theta = \frac{\Gamma \sqrt{\omega}}{\epsilon \sigma T^3_{ss}},
\end{equation}
where $\Gamma$ is the thermal inertia in
J~K$^{-1}$~m$^{-2}$~s$^{-1/2}$ and $\omega$ is the angular rotation
rate of the object in s$^{-1}$.  Slow rotators have $\Theta \ll 1$ and
fast rotators have $\Theta \gg 1$.  We chose 10 and
320~J~K$^{-1}$~m$^{-2}$~s$^{-1/2}$ as the thermal inertia extrema for
short period comets, as suggested by the observations of the sub-solar
temperature of Tempel~1 \citep{ahearn05} and ground-based observations
of (3200)~Phaethon \citep{green85}.  \citet{fernandez05} derived two
possible periods for the Encke nucleus, 11.1 or 22.2~hr.  The 11.1~hr
rotation period has been verified by radar observations
\citep{harmon05}.  To determine the sub-solar temperature we use our
best NEATM fits, $\eta = 1.026$ at 1.1~AU and $\eta = 0.735$ at
2.6~AU, which produces $T_{ss} = 382$~K at 1.1~AU and 269~K at 2.6~AU.
Together, $\Theta$ ranges 0.02--0.6 at 1.1~AU and 0.05--1.6 at 2.6~AU.
With these new measurements we conclude, as did \citep{fernandez05},
that at best we can use a slow rotator model (if $\Gamma \approx
10$~\mks) and at worst the Encke nucleus is in an intermediate state
between slow and fast rotation (if $\Gamma \approx 320$~\mks).
Additionally, there is no clear transition to a fast rotator as Encke
recedes from the sun.

We also note the recent mid-infrared measurements of asteroid
2002~NY40 by \citet{muller04} at phase angles of 22\degr{} and
59\degr{} show no $\eta$-phase angle dependence.  Although the
majority of the observations suggest the dependence \citep{wolters05},
the relation may not hold true for any particular object.  The case of
comet Encke is not resolved.  The observed SEDs are fit by an $\eta$
parameter that varies with phase angle but may also be fit by changing
the observed effective radius within the constraints of the derived
nucleus shape.  An SED of the Encke nucleus at high phase angle that
constrains both effective radius and color temperature is required to
support either possibility.

\subsection{Comet Mineralogy}\label{mineral-discuss}

The mineralogies of comet HT50's dust coma at both epochs are
consistent with each other given the errors and upper-limits
(Table~\ref{rel-mass}).  Also presented in Table~\ref{rel-mass} are
the mineralogies of comets Hale-Bopp and Tempel~1, pre- and
post-\textit{Deep Impact} encounter \citep{harker04, harker05}.  The
low mass ratio between the silicate minerals and the deeply absorbing
grains (represented by amorphous carbon) appears to be intermediate
between the aging Jupiter-family comet Encke and the presumably
pristine materials ejected from comet Hale-Bopp and comet Tempel~1
(post-\textit{Deep Impact}).  Does comet HT50 have a processed surface
similar to what may be on the surfaces of Jupiter-family comets?  Even
comet C/2001~Q4 (NEAT) exhibited a large range in silicate-to-carbon
ratios, varying from 2.7--5.7 in $\approx1$~hour \citep{wooden04}.
Comet Q4, like Hale-Bopp, showed strong jets \citep{lecacheux04} and
Q4's high silicate fractions may originate in the (local) jet
activity.  In comet jets, violent sublimation of ices and volatile
gases may excavate and entrain pristine grain materials from
sub-surface reservoirs within the nucleus, similar to the \textit{Deep
Impact} event.  Visual inspection of the \textit{Spitzer} images of
HT50 (Fig.~\ref{peakups}) show no distinct jet-features in the coma
and there were no reports of any outbursts by this comet.  Therefore,
it may be that comet HT50 is dominated by ``whole surface'' (global)
sublimation and that Jupiter-family and long period/Oort cloud comets
are somewhat similar in the structure of their immediate surfaces.
Indeed, it has been proposed that Oort cloud comets can form a
cohesive crust from normal re-surfacing processes, including galactic
cosmic rays \citep{strazzulla91}, supernovae \citep{stern88}, and
interstellar grain impacts \citep{stern86}.  It is apparent that more
work determining the mineralogy of comet comae is needed to understand
the extent of the silicate-to-carbon ratio variations in comets, its
correlation with nucleus activity, and possible connections to nucleus
surface structure.

\begin{figure}
\plottwo{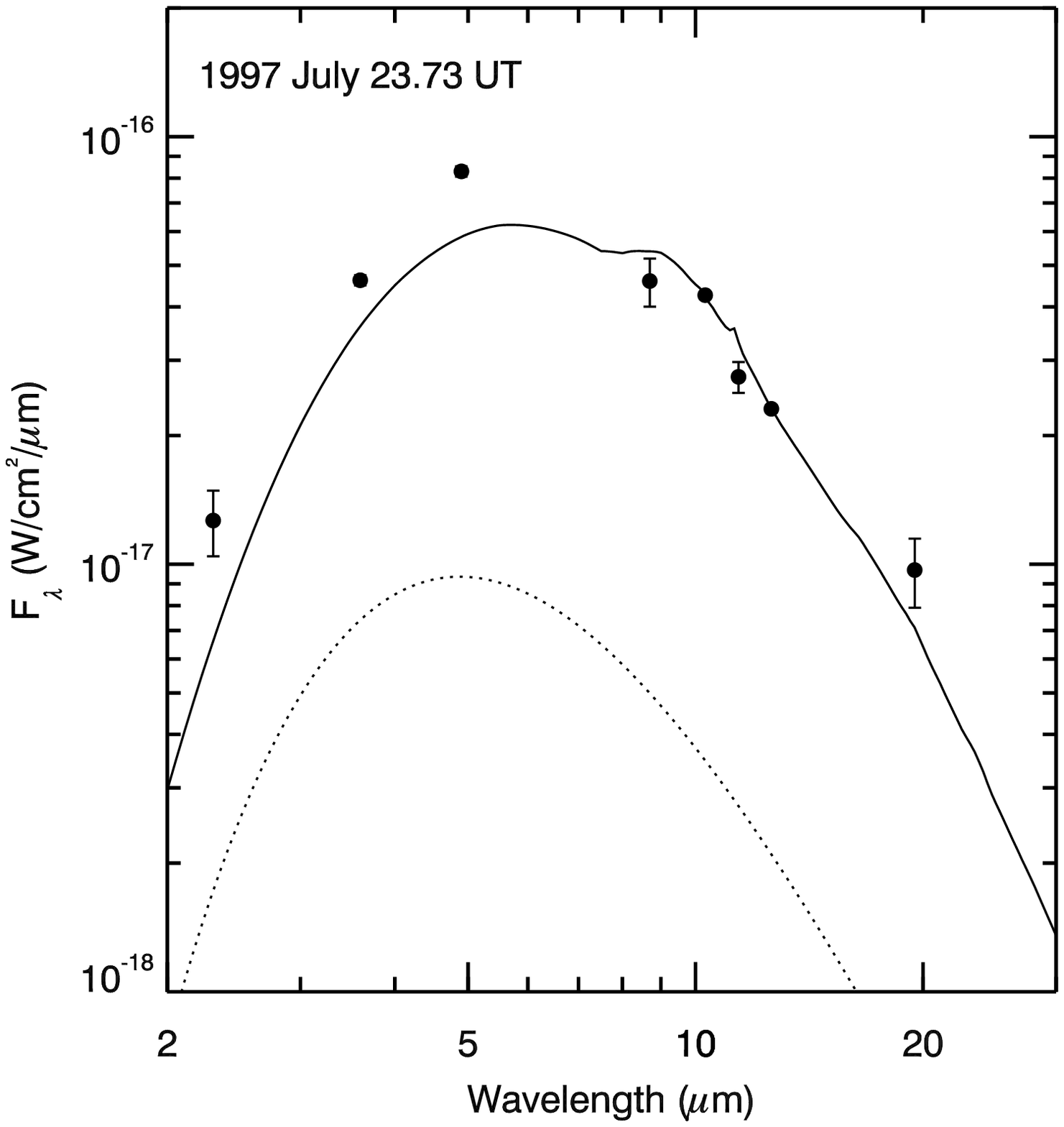}{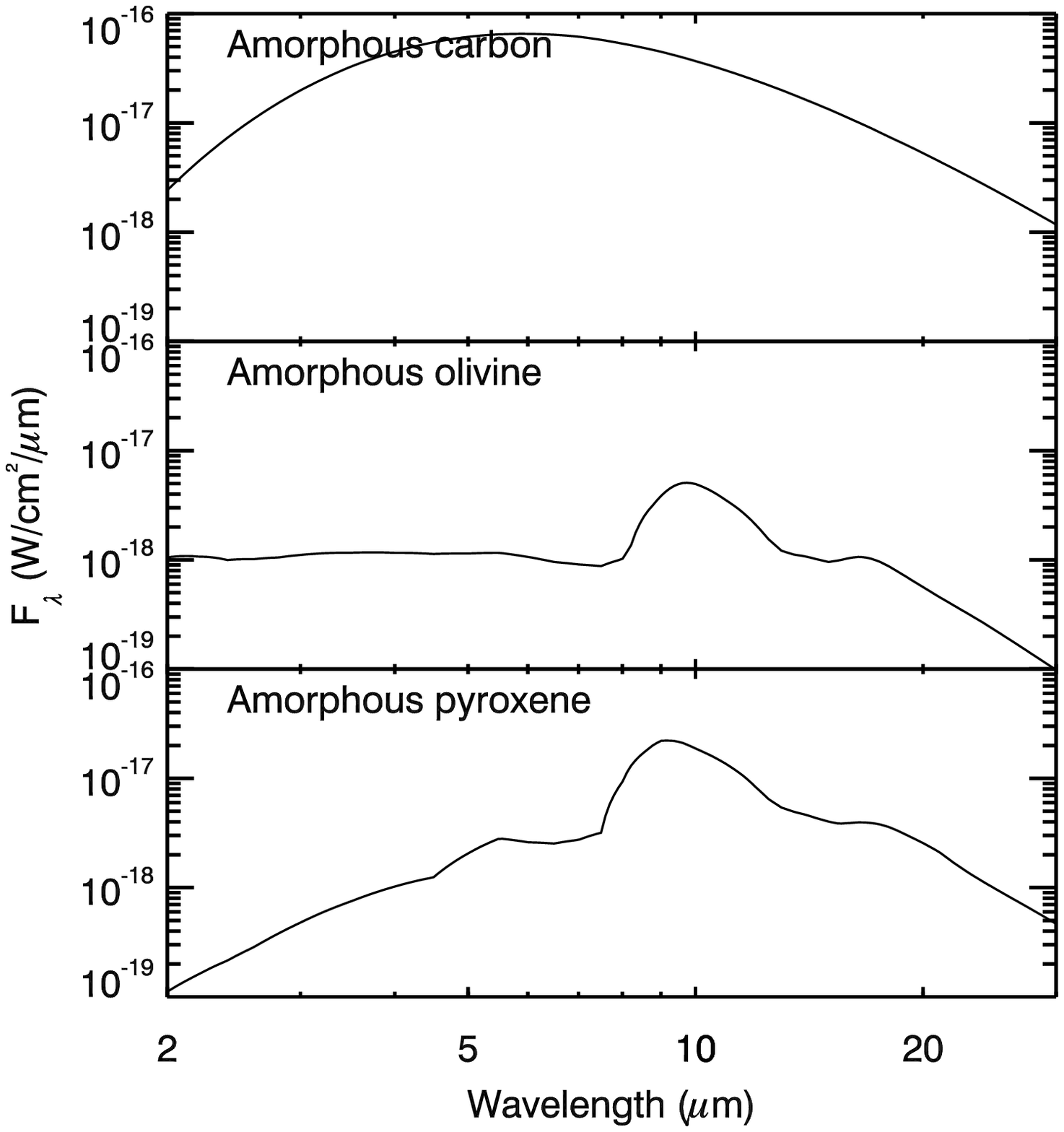}
\caption{Comet 2P/Encke fluxes from 1987 July \citep{gehrz89} and a
model calculated from the \textit{Spitzer}-derived coma mineralogy
upper-limits (solid-line; Table~\ref{encke-minerals}) and the
IRAC-derived nucleus parameters for $R=2.34$~km at a phase angle of
63\degr{} (dotted-line; Table~\ref{encke-fit-table}).  The models are
computed for the same viewing geometry as the photometer data ($r_h =
0.38$~AU, $\Delta=1.13$~AU, $\phi=63$\degr).  Also shown are the
decomposed spectra for each mineral except crystalline olivine, which
does not produce significant spectral features in the combined model
(see \S\ref{mineral-discuss}).}
\label{encke-jul87}
\end{figure}

Previous narrow-band photometry of comet Encke indicated a weak
silicate feature at small heliocentric distances \citep{campins82,
gehrz89} but no feature at 1.2~AU \citep{lisse04}.  The coma is
usually considered to be dominated by large particles, based upon a
low coma color-temperature and a lack of a strong silicate feature
\citep{gehrz89, reach00, lisse04}.  Instead, our best-fit models
indicate silicates are a minor constituent ($<31$\% by mass for $a
\leq 1$~\micron) to the coma.  To compare our results to previous
investigations near perihelion, we computed a ``silicate upper-limit''
model derived from our best-fit Encke mineralogy in
Table~\ref{encke-minerals}.  The upper-limit model sets the number of
peak grains for the silicate minerals to their three-sigma
upper-limits and decreases $N_p$ for the amorphous carbon component by
three-sigma.  The mineral ratio by sub-micron mass becomes 100:1:18:11
(amorphous carbon:amorphous olivine:amorphous pyroxene:crystalline
olivine).  The upper-limit model and photometry of comet Encke at
perihelion in 1987 July from \citet{gehrz89} is presented in
Fig.~\ref{encke-jul87}.  Also included is the Encke nucleus as derived
from Table~\ref{encke-fit-table} ($R=2.34$~km, $\phi=63$\degr).  The
coma and nucleus models were computed for the same geometry as the
photometry data ($r_h = 0.38$~AU, $\Delta=1.13$~AU, $\phi=63$\degr).
The coma model was scaled to account for Encke's difference in dust
production between $r_h=0.38$~AU and 2.4~AU.

The upper-limit model produces a silicate feature at 10~\micron{} that
approximates the weak silicate excess exhibited by the 9--13~\micron{}
photometry points.  The general shape of the SED is also approximated
by the model, except for the short wavelength flux points, which are
underestimated by factors of 1.2--1.8.  The model likely requires a
contribution from scattered sunlight to account for the 2.2~\micron{}
and 3.6~\micron{} data points.  Only the 5~\micron{} data point
remains unexplained (the error bar may not reflect variations in
atmospheric transparency through the optically thick 5~\micron{}
band-pass).  Modifying the relative contributions of the silicates and
carbonaceous minerals does not produce a satisfactory fit, although
the possibility remains that the PSD could be varied to account for
the spectral shape.  We conclude that our \textit{Spitzer}-derived
mineralogy is a robust estimate of 2P/Encke's coma composition and
that our best-fit model's PSD, derived at $r_h = 2.4$~AU, may not be
representative of the PSD observed throughout Encke's entire orbit.
Indeed, the brightness behavior near perihelion is asymmetric and is
best explained by different active regions on the nucleus, therefore,
a varying PSD about perihelion could be expected \citep{sekanina91}.

Jupiter-family comets are thought to form in the transneptunian region
and beyond \citep{duncan04, morbidelli04}.  We have derived
mineralogies for two Jupiter-family comets: comet Encke, likely
dominated by deeply absorbing grains, and comet Tempel~1 dominated by
silicate grains \citep{harker06}.  These are striking differences for
two comets assumed to arise from the same region of the proto-solar
disk.  The difference in mineralogy may be due to each comet's
original compositions or may be a reflection of different
compositional processing histories.  Ultraviolet and ion radiation
transforms relatively optically inactive organic material into a dark,
carbonized solid \citep{jenniskens93, greenberg95}.  Outgasing from
surfaces or sub-surface layers with different radiation exposures may
explain the differences in the coma mineralogy of comets Encke and
Tempel~1.  Alternatively, the outgasing surfaces or sub-surface layers
may have original and different compositions.  Such layering in
Jupiter-family comet nuclei has been proposed by \citet{belton06} to
account for topographical features on comets Tempel~1, Borrelly, and
Wild~2.

\section{CONCLUSIONS}
We present \textit{Spitzer}/IRS spectra of comets 2P/Encke,
67P/Churyumov-Gerasimenko, and C/2001~HT50 (LINEAR-NEAT), and
\textit{Spitzer}/IRAC and MIPS images of comet 2P/Encke.  Comet Encke
exhibited a smooth continuum, best modeled by carbonaceous grains with
a small peak grain size ($a_p=0.4$~\micron).  Previous investigations
into comet Encke's dust coma revealed a weak silicate feature at
perihelion ($r_h = 0.3$~AU).  We conclude the weak silicate feature is
due to the paucity of silicate grains and the preponderance of
carbonaceous grains (or some other warm, deeply absorbing material).
We constrain the sub-micron silicate fraction to $<31$\% by mass.  The
nucleus of comet Encke is fit by the near-Earth asteroid thermal model
with an effective radius $R = 2.34\pm0.14$.  The nucleus was observed
at phase angles 20\degr{} and 63\degr{} and may be exhibiting a
variation of the infrared beaming parameter with phase angle, which
indicates of a rough nucleus surface or appreciable night side
temperature.

Comet 67P exhibited a significant coma at a heliocentric distance of
5~AU, $F_\lambda = 2.01\pm0.10 \times 10^{-21}$~\wcm{} at
27.9~\micron{}.  67P's known dust trail comprises approximately 3\% of
the measured dust flux density.  The remaining coma flux was due to 1)
recently ejected dust (age of order hours to weeks), 2) large, slowly
moving particles from the 2003 perihelion passage, or 3) some
combination of the two.

Comet HT50 displayed a significant silicate mineralogy with a
silicate-to-carbon sub-micron mass ratio of 0.6.  The derived ratio of
0.6 is an order of magnitude lower than the silicate-to-carbon ratios
of post-\textit{Deep Impact} comet 9P/Tempel~1 and other Oort cloud
comets, C/1995~O1 (Hale-Bopp) and C/2001~Q4 (NEAT).

The differences in silicate-to-carbon mass ratios in comet comae may
be linked to strong jet activity in comets.  Comet HT50's derived
silicate-to-carbon sub-micron mass ratio is 0.6, but analysis of comet
Hale-Bopp, which exhibited strong jet activity, derived a ratio of
8.1.  At this time, the wide diversity in comet comae mineralogy
likely has not been probed.

\acknowledgements

The authors thank William Reach for useful discussions and comments.
In addition, the authors wish to thank the referee for their
insightful comments and suggestions that improved the manuscript.
This work is based on observations made with the Spitzer Space
Telescope, which is operated by the Jet Propulsion Laboratory,
California Institute of Technology under a contract with NASA.
Support for this work was provided by NASA through contracts 1256406,
1263741, 1275835, issued by JPL/Caltech to the University of
Minnesota.  C.E.W. and M.S.K. acknowledge support from the National
Science Foundation grant AST-037446.  M.S.K acknowledges support from
the University of Minnesota Doctoral Dissertation Fellowship.

Facilities: \facility{Spitzer}


\clearpage

\clearpage

\begin{deluxetable}{cccc}
\tablecolumns{4}
\tablewidth{0pt}

\tablecaption{Slit widths and sizes of the constant width extraction
apertures used for each module.\tablenotemark{a}\label{apertures}}

\tablehead{\colhead{IRS Module} &
  \colhead{Slit Width} &
  \colhead{Aperture Width} & 
  \colhead{Position Angle\tablenotemark{b}}\\
  &
  \colhead{(\arcsec)} &
  \colhead{(\arcsec)} &
  \colhead{(\degr)}}

\startdata
SL & 3.7  & 17.5 & 0 \\
SH & 4.7  & 11.3 & -43 \\
LL & 10.7 & 51.5 & -84 \\
LH & 11.1 & 22.3 & -128 \\
\enddata

\tablenotetext{a}{Adopted from the \textit{Spitzer} Observers' Manual
\citep{som} except SL and LL aperture widths (see
\S\ref{spectra-obs-text}).  SL = short-low, 5--14~\micron,
$R\approx64$--128; SH = short-high, 10--20~\micron, $R\approx600$; LL
= long-low, 14--38~\micron, $R\approx64$--128; LH = long-high,
20--38~\micron, $R\approx600$.}

\tablenotetext{b}{Relative to SL (values from SPOT,
\url{http://ssc.spitzer.caltech.edu/propkit/spot/}).}

\end{deluxetable}


\begin{deluxetable}{@{\hspace{10mm}}lllcllc}
\tablecolumns{7}
\tablecaption{Summary of comet spectroscopy.\label{spec-log}}

\tablehead{\colhead{Comet / Date} &
  \colhead{PID\tablenotemark{a}} &
  \colhead{IRS Modules} &
  \colhead{Pre/Post-} &
  \colhead{$r_h$\tablenotemark{b}} &
  \colhead{$\Delta_s$\tablenotemark{c}} &
  \colhead{Phase}\\
  \colhead{(UT)} & & &
  \colhead{Perihelion} &
  \colhead{(AU)} &
  \colhead{(AU)} &
  \colhead{Angle\tablenotemark{d} (\degr)} }

\startdata
\multicolumn{7}{l}{2P/Encke} \\
2004 Jun 25.23 & 210 & Red, SL, SH, LH & Post & 2.573 & 1.985 & 21 \\
2004 Jun 25.76 & 119 & Blue, SL, LL    & Post & 2.577 & 1.982 & 21 \\
\multicolumn{7}{l}{67P/Churyumov-Gerasimenko} \\
2004 Jul 15.34 & 2316 & Red, LL, LH & Post & 4.978 & 4.743 & 12 \\
\multicolumn{7}{l}{C/2001 HT50 (LINEAR-NEAT)} \\
2003 Dec 17.65 & 131 & Red, SL, SH, LH & Post & 3.238 & 2.652 & 16 \\
2004 Jul 18.44 & 131 & Red, SL, SH, LH & Post & 4.598 & 4.368 & 13 \\
\enddata

\tablenotetext{a}{\textit{Spitzer} Program ID}

\tablenotetext{b}{Red = 22~\micron{} peak-up; Blue = 16~\micron{}
peak-up; SL = short-low, 5--14~\micron, $R\approx64$--128; SH =
short-high, 10--20~\micron, $R\approx600$; LL = long-low,
14--38~\micron, $R\approx64$--128; LH = long-high, 20--38~\micron,
$R\approx600$.}

\tablenotetext{c}{\textit{Spitzer}-comet distance.}

\tablenotetext{d}{Sun-comet-\textit{Spitzer} angle.}

\end{deluxetable}

\begin{deluxetable}{lr@{.}lr@{.}lr@{.}lr@{.}l}
\tablecolumns{5}
\tablewidth{0pt}

\tablecaption{The parameters of third-order polynomial fits to the
slit-loss correction factors versus wavelength (in units of
\micron).\tablenotemark{a}\label{slcf-table}}

\tablehead{\colhead{Module} & \multicolumn{2}{c}{$\lambda_0$} &
 \multicolumn{2}{c}{$\lambda_1$} & \multicolumn{2}{c}{$\lambda_2$} &
 \multicolumn{2}{c}{$\lambda_3$}}

\startdata
SL2 &      1&8591233 &      -0&3691497 &       0&0425228 &      -0&0017076 \\
SL1 &      0&8424725 &      -0&0064947 &      -0&0006418 &       0&0000008 \\
SH  &      0&6111256 &       0&0466900 &      -0&0040512 &       0&0000857 \\
LL2 &      1&9293931 &      -0&1338932 &       0&0052684 &      -0&0000686 \\
LL1 &      0&8016602 &       0&0051615 &      -0&0002820 &       0&0000017 \\
LH  &      1&0151328 &      -0&0165096 &       0&0004606 &      -0&0000064 \\
\enddata

\tablenotetext{a}{The slit-loss correction factor spectrally
calibrates extended sources of uniform surface brightness (see
\S\ref{spectra-obs-text} and Fig.~\ref{slcf}).}

\end{deluxetable}

\begin{deluxetable}{llcllc}
\tablecolumns{6}
\tablewidth{0pt}
\tabletypesize{\footnotesize}
\tablecaption{Summary of comet 2P/Encke imagery.\label{image-log}}

\tablehead{\colhead{Date} & \colhead{Wavelength} & \colhead{Pre/Post-}
        & \colhead{r} & \colhead{$\Delta$} & \colhead{Phase} \\
        \colhead{(UT)} & \colhead{(\micron)} & \colhead{Perihelion} &
        \colhead{(AU)} & \colhead{(AU)} & \colhead{Angle (\degr)}}

\startdata
2003 Nov 11.73 & 3.6, 4.5, 5.8    & Pre & 1.094 & 0.232 & 63 \\
2004 Jun 23.21 & 24, 70           & Post & 2.556 & 1.997 & 22 \\
2004 Jun 29.92 & 3.6, 4.5, 5.8, 8 & Post & 2.611 & 1.958 & 20 \\
\enddata

\end{deluxetable}

\begin{deluxetable}{lcr@{ $\pm$ }l@{ $\times$}l}
\tablecolumns{4}
\tablewidth{0pt}
\tablecaption{2P/Encke color and aperture corrected nucleus fluxes.\label{encke-nuke}}

\tablehead{\colhead{Date} &
  \colhead{Wavelength\tablenotemark{a}} &
  \multicolumn{3}{c}{Flux} \\
  \colhead{(UT)} &
  \colhead{(\micron)} &
  \multicolumn{3}{c}{(\wcm)}}

\startdata
  2003 Nov 17.75 & 3.55 & 1.986 & 0.061 & $10^{-18}$ \\
                 & 4.49 & 5.06 & 0.11 & $10^{-18}$ \\
		 & 5.73 & 1.007 & 0.021 & $10^{-17}$ \\
  2004 Jun 23.21 & 23.7 & 2.30 & 0.24 & $10^{-20}$ \\
                 & 71.4 & 1.27 & 0.35 & $10^{-21}$ \\
  2004 Jun 29.92 & 3.55 & 7.0 & 2.8 & $10^{-22}$ \\
                 & 4.49 & 3.95 & 0.42 & $10^{-21}$ \\
                 & 5.73 & 1.07 & 0.11 & $10^{-20}$ \\
                 & 7.87 & 3.99 & 0.12 & $10^{-20}$ \\
\enddata

\tablenotetext{a}{IRAC or MIPS effective wavelengths.}

\end{deluxetable}

\begin{deluxetable}{lr@{ $\pm$ }lr@{ $\pm$ }lr@{ $\pm$ }l}
\tablecolumns{7}
\tablewidth{0pt}
\tablecaption{2P/Encke NEATM fit to the 2004 June and 2003 November
data\tablenotemark{a}.\label{encke-fit-table}}

\tablehead{\colhead{Parameter} &
	\multicolumn{2}{c}{2004 June} &
	\multicolumn{4}{c}{2003 November} \\
	& \multicolumn{2}{c}{} &
	\multicolumn{2}{c}{Fixed $\eta$} &
	\multicolumn{2}{c}{Fixed $R$}}

\startdata
radius, $R$ (km) & 2.34 & 0.14 &
	1.72 & 0.10 &
	\multicolumn{2}{c}{2.34} \\

geometric albedo, $p_v$ & \multicolumn{2}{c}{0.047} &
	\multicolumn{2}{c}{0.047} &
	\multicolumn{2}{c}{0.047} \\

IR beaming parameter, $\eta$ & 0.735 & 0.046 &
	\multicolumn{2}{c}{0.735} &
	1.026 & 0.061 \\

IR emissivity, $\epsilon$ & \multicolumn{2}{c}{0.9} &
	\multicolumn{2}{c}{0.9} &
	\multicolumn{2}{c}{0.9} \\

reflected light scale, $\alpha$ & 0.047 & 0.027 &
	0.031 & 0.022 &
	0.127 & 0.016 \\

degrees of freedom, $\nu$ & \multicolumn{2}{c}{3} &
	\multicolumn{2}{c}{1} &
	\multicolumn{2}{c}{1} \\

$\chi^2_\nu$ & \multicolumn{2}{c}{3.5} &
	\multicolumn{2}{c}{11.6} &
	\multicolumn{2}{c}{0.11} \\
\enddata

\tablenotetext{a}{Values without error bars are fixed parameters.}

\end{deluxetable}

\begin{deluxetable}{lllll}
\tablecolumns{5}
\tablewidth{0pt}

\tablecaption{Scaling factors for each extracted
spectrum.\tablenotemark{a}\label{scales}}

\tablehead{\colhead{Comet} & \colhead{SL} & \colhead{SH} &
\colhead{LL} & \colhead{LH}}

\startdata
2P/Encke (SL+LL) & 1.53 & \nodata & 1.30 & \nodata \\
2P/Encke (SH+LH) & \nodata & 0.48 & \nodata & 0.58 \\
C/2001~HT50 (LINEAR-NEAT) (3.2~AU) & 1.24 & 1.00 & \nodata & 1.51 \\
C/2001~HT50 (LINEAR-NEAT) (4.6~AU) & 1.81 & 1.00 & \nodata & 1.23 \\
\enddata

\tablenotetext{a}{The spectrum of comet 67P/Churyumov-Gerasimenko was
not scaled.}

\end{deluxetable}


\begin{deluxetable}{lllllr@{ $\pm$ }lr@{ $\pm$ }lr@{ $\pm$ }lr@{ $\pm$ }lll}
\tablecolumns{15}
\tablewidth{0pt}
\rotate

\tablecaption{Comet 2P/Encke best-fit thermal emission model
parameters.\label{encke-minerals}}

\tablehead{&&&&&\multicolumn{8}{c}{$N_p$ ($\times 10^{18}$)} \\
\cline{6-13} \colhead{Data} & \colhead{$N$} & \colhead{$M$} &
\colhead{$a_p$\tablenotemark{a}} & \colhead{$D$\tablenotemark{b}} &
\multicolumn{2}{c}{Amorphous} & \multicolumn{2}{c}{Amorphous} &
\multicolumn{2}{c}{Amorphous} & \multicolumn{2}{c}{Crystalline} &
\colhead{$\nu$} & \colhead{$\chi^2_\nu$} \\ & & & \colhead{(\micron)}
& & \multicolumn{2}{c}{Carbon\tablenotemark{c}} &
\multicolumn{2}{c}{Olivine} & \multicolumn{2}{c}{Pyroxene} &
\multicolumn{2}{c}{Olivine}}

\startdata
SL+LL & 3.7 & 11.1 & 0.4 & 2.857 &
  2.38 & $^{0.12}_{0.17}$ & \multicolumn{2}{c}{$<$0.14} &
  \multicolumn{2}{c}{$<$0.77} & \multicolumn{2}{c}{$<$0.07} &
  240 & 0.97 \\
SH+LH & 3.7 & 11.1 & 0.4 & 2.857 &
  2.54 & $^{0.06}_{0.17}$ & \multicolumn{2}{c}{$<$0.03} &
  \multicolumn{2}{c}{$<$0.44} & \multicolumn{2}{c}{$<$0.10} &
  1473 & 0.47 \\
\enddata

\tablenotetext{a}{Our estimated error in $a_p$ is 0.05~\micron{} (see
\S\ref{encke-coma}).}

\tablenotetext{b}{Fractal dimension of the amorphous components;
crystals always have a fractal dimension of 3.0 (i.e., solid grains).}

\tablenotetext{c}{Amorphous carbon represents the warm, deeply
absorbing component in comet dust.}

\end{deluxetable}


\begin{deluxetable}{llllllr@{ $\pm$ }lr@{ $\pm$ }lr@{ $\pm$ }lr@{ $\pm$ }lll}
\tablecolumns{16}
\tablewidth{0pt}
\rotate

\tablecaption{Comet C/2001~HT50 (LINEAR-NEAT) best-fit thermal
emission model parameters.\label{ht50-minerals}}

\tablehead{&&&&&&\multicolumn{8}{c}{$N_p$ ($\times 10^{19}$)} \\
\cline{7-14} \colhead{Date} & \colhead{$r$} & \colhead{$N$} &
\colhead{$M$} & \colhead{$a_p$\tablenotemark{a}} & \colhead{$D$} &
\multicolumn{2}{c}{Amorphous} & \multicolumn{2}{c}{Amorphous} &
\multicolumn{2}{c}{Amorphous} & \multicolumn{2}{c}{Crystalline} &
\colhead{$\nu$} & \colhead{$\chi^2_\nu$} \\ \colhead{(UT)} &
\colhead{(AU)} & & & \colhead{(\micron)} & &
\multicolumn{2}{c}{Carbon} & \multicolumn{2}{c}{Olivine} &
\multicolumn{2}{c}{Pyroxene} & \multicolumn{2}{c}{Olivine}}

\startdata
2003 Dec 17.65 & 3.2 & 4.2 & 46.2 & 1.2 & 3.000 &
  3.12 & $^{0.03}_{0.13}$ & 0.12 & $^{0.10}_{0.03}$ &
  \multicolumn{2}{c}{$<$0.03} & 1.27 & $^{0.86}_{0.32}$ &
  1686 & 4.2 \\
2004 Jul 18.44 & 4.6 & 3.7 & 29.6 & 1.2 & 2.857 &
  2.98 & $^{0.61}_{0.03}$ & 0.49 & $^{0.73}_{0.09}$ &
  \multicolumn{2}{c}{$<$0.85} &
  \multicolumn{2}{c}{$<$3.4} & 1690 & 23.8 \\
\enddata

\tablenotetext{a}{Our estimated error in $a_p$ is 0.05~\micron{} (see
\S\ref{encke-coma}).}

\end{deluxetable}


\begin{deluxetable}{lr@{.}lr@{.}l@{ $\pm$ }lr@{.}l@{ $\pm$ }lr@{.}l@{ $\pm$ }lr@{.}l@{ $\pm$ }lr@{.}l@{ $\pm$ }l}
\rotate
\tablecolumns{18}
\tabletypesize{\small}
\tablewidth{0pt}

\tablecaption{Relative mineralogy of comets 2P/Encke and C/2001~HT50
(LINEAR-NEAT) by mass of sub-micron grains derived from
Tables~\ref{encke-minerals} and
\ref{ht50-minerals}\tablenotemark{a}.\label{rel-mass}}

\tablehead{\colhead{Comet} & \multicolumn{2}{c}{$r$} &
\multicolumn{3}{c}{Amorphous} & \multicolumn{3}{c}{Amorphous} &
\multicolumn{3}{c}{Amorphous} & \multicolumn{3}{c}{Crystalline} &
\multicolumn{3}{c}{Silicate /} \\ &\multicolumn{2}{c}{(AU)} &
\multicolumn{3}{c}{Carbon} & \multicolumn{3}{c}{Olivine} &
\multicolumn{3}{c}{Pyroxene} & \multicolumn{3}{c}{Olivine} &
\multicolumn{3}{c}{Carbon}}

\startdata

2P/Encke\tablenotemark{b} & 2&6 & 1&00 & $^{0.00}_{0.07}$
                               & $<$0&\multicolumn{2}{@{}l}{01}
                               & $<$0&\multicolumn{2}{@{}l}{14}
                               & $<$0&\multicolumn{2}{@{}l}{05}
                               & $<$0&\multicolumn{2}{@{}l}{08} \\
C/2001~HT50 (LINEAR-NEAT) & 3&2 & 0&63 & $^{0.01}_{0.05}$
             & 0&032 & $^{0.012}_{0.004}$
             & $<$0&\multicolumn{2}{@{}l}{006}
             & 0&34 & $^{0.05}_{0.02}$
             & 0&59 & $^{0.14}_{0.04}$ \\
C/2001~HT50 (LINEAR-NEAT) & 4&6 & 0&82 & $^{0.04}_{0.25}$
             & 0&18 & $^{0.11}_{0.08}$
             & $<$0&\multicolumn{2}{@{}l}{14}
             & $<$0&\multicolumn{2}{@{}l}{60}
             & $<$2&\multicolumn{2}{@{}l}{2} \\
C/1995~O1 (Hale-Bopp)\tablenotemark{c} & 2&8 & 0&\multicolumn{2}{@{}l}{11}
             & 0&\multicolumn{2}{@{}l}{18}
             & 0&\multicolumn{2}{@{}l}{18}
             & 0&\multicolumn{2}{@{}l}{48}
             & 8&\multicolumn{2}{@{}l}{1} \\
9P/Tempel~1 (pre-\textit{DI}) & 1&5 & 0&\multicolumn{2}{@{}l}{0}
             & 1&\multicolumn{2}{@{}l}{0}
             & 0&\multicolumn{2}{@{}l}{0}
             & 0&\multicolumn{2}{@{}l}{0}
             & \multicolumn{3}{l}{\phn\nodata} \\
9P/Tempel~1 (\textit{DI}+1.0)\tablenotemark{d} & 1&5 & 0&\multicolumn{2}{@{}l}{21}
             & 0&\multicolumn{2}{@{}l}{27}
             & 0&\multicolumn{2}{@{}l}{41}
             & 0&\multicolumn{2}{@{}l}{10}
             & 3&\multicolumn{2}{@{}l}{7} \\
\enddata

\tablenotetext{a}{Included for comparison are comets C/1995~O1
(Hale-Bopp) and 9P/Tempel~1 pre- and post-\textit{Deep Impact} encounter
\citep{harker04,harker05}. }

\tablenotetext{b}{Values derived from the SH+LH model fit of
Table~\ref{encke-minerals}.}

\tablenotetext{c}{The remaining 5\% is composed of orthopyroxene.}

\tablenotetext{d}{Time of impact +1.0 hour.}

\end{deluxetable}


\begin{thebibliography}{}
\bibitem[A'Hearn(2004)]{ahearn04} A'Hearn, M.~F.\ 2004, Comets II, 17

\bibitem[A'Hearn et al.(2005)]{ahearn05} A'Hearn, M.~F., et al.\ 2005,
Science, 310, 258

\bibitem[Belton \& Deep Impact Science Team(2006)]{belton06} Belton,
M.~J.~S., \& Deep Impact Science Team 2006, 37th Annual Lunar and
Planetary Science Conference, 37, 1232

\bibitem[Brownlee et al.(2004)]{brownlee04} Brownlee, D.~E., et al.\
2004, Science, 304, 1764

\bibitem[Campins(1988)]{campins88} Campins, H.\ 1988, Icarus, 73, 508

\bibitem[Campins et al.(1987)]{campins87} Campins, H., A'Hearn, M.~F.,
\& McFadden, L.-A.\ 1987, \apj, 316, 847

\bibitem[Campins et al.(1982)]{campins82} Campins, H., Rieke, G.~H.,
\& Lebofsky, M.~J.\ 1982, Icarus, 51, 461

\bibitem[Delb{\'o} \& Harris(2002)]{delbo02} Delb{\'o}, M., \& Harris,
A.~W.\ 2002, Meteoritics and Planetary Science, 37, 1929
 
\bibitem[Delb{\'o} et al.(2003)]{delbo03} Delb{\'o}, M., Harris,
A.~W., Binzel, R.~P., Pravec, P., \& Davies, J.~K.\ 2003, Icarus, 166,
116

\bibitem[Dones et al.(2004)]{dones04} Dones, L., Weissman, P.~R.,
Levison, H.~F., \& Duncan, M.~J.\ 2004, Comets II, 153

\bibitem[Duncan et al.(2004)]{duncan04} Duncan, M., Levison, H., \&
Dones, L.\ 2004, Comets II, 193

\bibitem[Ehrenfreund et al.(2004)]{ehrenf04} Ehrenfreund, P.,
Charnley, S.~B., \& Wooden, D.\ 2004, Comets II, 115

\bibitem[Fazio et al.(2004)]{fazio04} Fazio, G.~G., et al.\ 2004,
\apjs, 154, 10
 
\bibitem[Fernandez(1999)]{fernandez99} Fernandez, Y.~R.\ 1999,
Ph.D.~Thesis (University of Maryland, College Park)

\bibitem[Fern{\'a}ndez et al.(2005)]{fernandez05} Fern{\'a}ndez,
Y.~R., Lowry, S.~C., Weissman, P.~R., Mueller, B.~E.~A., Samarasinha,
N.~H., Belton, M.~J.~S., \& Meech, K.~J.\ 2005, Icarus, 175, 194

\bibitem[Fern{\'a}ndez et al.(2000)]{fernandez00} Fern{\'a}ndez,
Y.~R., et al.\ 2000, Icarus, 147, 145

\bibitem[Gehrz et al.(1989)]{gehrz89} Gehrz, R.~D., Ney, E.~P.,
Piscitelli, J., Rosenthal, E., \& Tokunaga, A.~T.\ 1989, Icarus, 80,
280

\bibitem[Gehrz et al.(2006)]{gehrz06} Gehrz, R.~D., Reach, W.~T.,
Woodward, C.~E., \& Kelley, M.~S.\ 2006, Adv.\ Space\ Res., in press

\bibitem[Green et al.(1985)]{green85} Green, S.~F., Meadows, A.~J., \&
Davies, J.~K.\ 1985, \mnras, 214, 29P

\bibitem[Greenberg et al.(1995)]{greenberg95} Greenberg, J.~M., Li,
A., Mendoza-Gomez, C.~X., Schutte, W.~A., Gerakines, P.~A., \& de
Groot, M.\ 1995, \apjl, 455, L177

\bibitem[Hanner(1983)]{hanner83} Hanner, M.~S.\ 1983, Cometary
exploration; Proceedings of the International Conference, Budapest,
Hungary, November 15-19, 1982.~Volume 2 (A84-47701 23-90).~Budapest,
Akademiai Kiado, 1983, p.~1-22.~NASA-supported research., 2, 1

\bibitem[Hanner et al.(1981)]{hanner81} Hanner, M.~S., Giese, R.~H.,
Weiss, K., \& Zerull, R.\ 1981, \aap, 104, 42

\bibitem[Hanner et al.(1996)]{hanner96} Hanner, M.~S., Lynch, D.~K.,
Russell, R.~W., Hackwell, J.~A., Kellogg, R., \& Blaney, D.\ 1996,
Icarus, 124, 344

\bibitem[Harker(1999)]{harker99a} Harker, D.~E.\ 1999, Ph.D.~Thesis
(Univ. Wyoming)

\bibitem[Harker et al.(2002)]{harker02} Harker, D.~E., Wooden, D.~H.,
Woodward, C.~E., \& Lisse, C.~M.\ 2002, \apj, 580, 579

\bibitem[Harker et al.(2004)]{harker04} Harker, D.~E., Wooden, D.~H.,
Woodward, C.~E., \& Lisse, C.~M.\ 2004, \apj, 615, 1081

\bibitem[Harker et al.(2005)]{harker05} Harker, D.~E., Woodward,
C.~E., \& Wooden, D.~H.\ 2005, Science, 310, 278

\bibitem[Harker et al.(2006)]{harker06} Harker, D.~E., Woodward,
C.~E., Wooden, D.~H., Fisher, S., \& Trujullio, C. \ 2006, Icarus,
submitted

\bibitem[Harker et al.(1999)]{harker99} Harker, D.~E., Woodward,
C.~E., Wooden, D.~H., Witteborn, F.~C., \& Meyer, A.~W.\ 1999, \aj,
118, 1423

\bibitem[Harmon \& Nolan(2005)]{harmon05} Harmon, J.~K., \& Nolan,
M.~C.\ 2005, Icarus, 176, 175

\bibitem[Harris(1998)]{harris98} Harris, A.~W.\ 1998, Icarus, 131, 291

\bibitem[Honda et al.(2004)]{honda04} Honda, M., et al.\ 2004, \apj,
601, 577

\bibitem[Houck et al.(2004)]{houck04} Houck, J.~R., et al.\ 2004,
\apjs, 154, 18

\bibitem[Jenniskens et al.(1993)]{jenniskens93} Jenniskens, P.,
Baratta, G.~A., Kouchi, A., de Groot, M.~S., Greenberg, J.~M., \&
Strazzulla, G.\ 1993, \aap, 273, 583

\bibitem[Jewitt(1991)]{jewitt91} Jewitt, D.\ 1991, ASSL Vol.~167: IAU
Colloq.~116: Comets in the post-Halley era, 19

\bibitem[Keller et al.(1986)]{keller86} Keller, H.~U., et al.\ 1986,
\nat, 321, 320

\bibitem[Kelley et al.(2005a)]{kelley05} Kelley, M.~S., Reach, W.~T.,
\& Lien, D.~J.\ 2005a, Icarus, submitted

\bibitem[Kelley et al.(2005b)]{kelley05b} Kelley, M.~S., et al.\
2005b, American Astronomical Society Meeting Abstracts, 206, 34.03

\bibitem[Lagerros(1998)]{lagerros98} Lagerros, J.~S.~V.\ 1998, \aap,
332, 1123

\bibitem[Lamy et al.(2004)]{lamy04} Lamy, P.~L., Toth, I., Fernandez,
Y.~R., \& Weaver, H.~A.\ 2004, Comets II, 223

\bibitem[Lamy et al.(2003)]{lamy03} Lamy, P.~L., Toth, I., Weaver, H.,
Jorda, L., \& Kaasalainen, M.\ 2003, AAS/Division for Planetary
Sciences Meeting Abstracts, 35, 30.04

\bibitem[Lebofsky et al.(1986)]{lebofsky86} Lebofsky, L.~A., et al.\
1986, Icarus, 68, 239

\bibitem[Lecacheux \& Frappa(2004)]{lecacheux04} Lecacheux, J., \&
Frappa, E.\ 2004, \iaucirc, 8349, 1

\bibitem[Lisse et al.(2004)]{lisse04} Lisse, C.~M., et al.\ 
2004, Icarus, 171, 444

\bibitem[Lisse et al.(2005)]{lisse05} Lisse, C.~M., et al.\ 2005,
\apjl, 625, L139

\bibitem[Lynch et al.(2000)]{lynch00} Lynch, D.~K., Russell, R.~W., \&
Sitko, M.~L.\ 2000, Icarus, 144, 187

\bibitem[Lynch et al.(2002)]{lynch02} Lynch, D.~K., Russell, R.~W., \&
Sitko, M.~L.\ 2002, Icarus, 159, 234

\bibitem[Makovoz \& Khan(2005)]{makovoz05} Makovoz, D., Khan, I.\
2005.\ Mosaicking with MOPEX.\ Astronomical Society of the Pacific
Conference Series 347, 81.

\bibitem[Morbidelli \& Brown(2004)]{morbidelli04} Morbidelli, A., \&
Brown, M.~E.\ 2004, Comets II, 175

\bibitem[M{\"u}ller et al.(2004)]{muller04} M{\"u}ller, T.~G.,
Sterzik, M.~F., Sch{\"u}tz, O., Pravec, P., \& Siebenmorgen, R.\ 2004,
\aap, 424, 1075

\bibitem[Petrova et al.(2001)]{petrova01} Petrova, E.~V., Jockers, K.,
\& Kiselev, N.~N.\ 2001, Solar System Research, 35, 390

\bibitem[Pravdo et al.(2001)]{pravdo01} Pravdo, S., Helin, E.,
Lawrence, K., \& Spahr, T.~B.\ 2001, \iaucirc, 7624, 1

\bibitem[Reach et al.(2000)]{reach00} Reach, W.~T., Sykes, M.~V.,
Lien, D., \& Davies, J.~K.\ 2000, Icarus, 148, 80

\bibitem[Reach et al.(2005)]{reach05} Reach, W.~T., et al.\ 2005,
\pasp, 117, 978

\bibitem[Rieke et al.(2004)]{rieke04} Rieke, G.~H., et al.\ 2004,
\apjs, 154, 25

\bibitem[Schleicher(2006)]{schleicher06} Schleicher, D.~G.\ 2006,
Icarus, 181, 442

\bibitem[Sekanina(1991)]{sekanina91} Sekanina, Z.\ 1991, \jrasc, 85,
324

\bibitem[Shkuratov et al.(2000)]{shkuratov00} Shkuratov, Y.,
Stankevich, D., Sitko, M.~L., \& Sprague, A.~L.\ 2000, ASP
Conf.~Ser.~196: Thermal Emission Spectroscopy and Analysis of Dust,
Disks, and Regoliths, 196, 221

\bibitem[Sitko et al.(2004)]{sitko04} Sitko, M.~L., Lynch, D.~K.,
Russell, R.~W., \& Hanner, M.~S.\ 2004, \apj, 612, 576

\bibitem[Soderblom et al.(2002)]{soderblom02} Soderblom, L.~A., et
al.\ 2002, Science, 296, 1087

\bibitem[Spencer et al.(1989)]{spencer89} Spencer, J.~R., Lebofsky,
L.~A., \& Sykes, M.~V.\ 1989, Icarus, 78, 337

\bibitem[Spitzer Science Center(2005)]{som} Spitzer Science Center.\
2005, Spitzer Observers' Manual (Pasadena: SSC),
http://ssc.spitzer.caltech.edu/documents/som/

\bibitem[Spitzer Science Center(2006a)]{idh} Spitzer Science Center.\
2006a, Infrared Array Camera Data Handbook (Pasadena: SSC),
http://ssc.spitzer.caltech.edu/irac/dh/

\bibitem[Spitzer Science Center(2006b)]{mdh} Spitzer Science Center.\
2006b, Multiband Imaging Photometer for Spitzer (MIPS) Data Handbook
(Pasadena: SSC), http://ssc.spitzer.caltech.edu/mips/dh/

\bibitem[Stansberry et al.(2004)]{stansberry04} Stansberry, J.~A., et
al.\ 2004, \apjs, 154, 463

\bibitem[Strazzulla \& Johnson(1991)]{strazzulla91} Strazzulla, G., \&
Johnson, R.~E.\ 1991, ASSL Vol.~167: IAU Colloq.~116: Comets in the
post-Halley era, 243

\bibitem[Stern(1986)]{stern86} Stern, S.~A.\ 1986, Icarus, 68, 276

\bibitem[Stern(2003)]{stern03} Stern, S.~A.\ 2003, \nat, 424, 639

\bibitem[Stern \& Shull(1988)]{stern88} Stern, S.~A., \& Shull, J.~M.\
1988, \nat, 332, 407

\bibitem[Sykes \& Walker(1992)]{sykes92} Sykes, M.~V., \& Walker,
R.~G.\ 1992, Icarus, 95, 180
 
\bibitem[Veeder et al.(1987)]{veeder87} Veeder, G.~J., Hanner, M.~S.,
\& Tholen, D.~J.\ 1987, \aj, 94, 169

\bibitem[Werner et al.(2004)]{werner04} Werner, M.~W., et al.\ 
2004, \apjs, 154, 1

\bibitem[Whipple \& Hamid(1950)]{whipple50} Whipple, F.~L., \& Hamid,
S.~E.-D.\ 1950, \aj, 55, 185

\bibitem[Wolters et al.(2005)]{wolters05} Wolters, S.~D., Green,
S.~F., McBride, N., \& Davies, J.~K.\ 2005, Icarus, 175, 92

\bibitem[Wooden et al.(2005)]{wooden05} Wooden, D.~H., Harker, D.~E.,
\& Brearley, A.~J.\ 2005, ASP Conf.~Ser.~341: Chondrites and the
Protoplanetary Disk, 341, 774

\bibitem[Wooden et al.(2004)]{wooden04} Wooden, D.~H., Woodward,
C.~E., \& Harker, D.~E.\ 2004, \apjl, 612, L77

\end{thebibliography}
\end{document}